\documentclass[pra,showpacs]{revtex4-1}
\usepackage{color}
\usepackage{graphicx}
\usepackage[center]{subfigure}

\begin{document}

\title{Pattern formation by kicked solitons in the two-dimensional
Ginzburg-Landau medium with a transverse grating}
\author{Valentin Besse$^{1}$, Herv\'{e} Leblond$^{1}$, Dumitru Mihalache$%
^{1,2,3}$ and Boris A. Malomed$^{4}$}

\begin{abstract}
We consider the kick (tilt)-induced mobility of two-dimensional (2D)
fundamental dissipative solitons in models of bulk lasing media based on the
2D complex Ginzburg-Landau (CGL) equation including a spatially periodic
potential (transverse grating). The depinning threshold, which depends on
the orientation of the kick, is identified by means of systematic
simulations, and estimated by means of an analytical approximation. Various
pattern-formation scenarios are found above the threshold. Most typically,
the soliton, hopping between potential cells, leaves arrayed patterns of
different sizes in its wake. In the single-pass-amplifier setup, this effect
may be used as a mechanism for the selective pattern formation controlled by
the tilt of the input beam. Freely moving solitons feature two distinct
values of the established velocity. Elastic and inelastic collisions between
free solitons and pinned arrayed patterns are studied too.
\end{abstract}

\pacs{42.65.Tg,42.65.Sf,47.20.Ky}

\affiliation{$^{1}$ LUNAM Universit\'{e}, Universit\'{e} d'Angers,
Laboratoire de Photonique d'Angers, EA 4464,
2 Bd Lavoisier, 49000 Angers, France\\
$^{2}$ Horia Hulubei National Institute for Physics and Nuclear Engineering,
30 Reactorului, Magurele-Bucharest, 077125, Romania\\
$^{3}$ Academy of Romanian Scientists, 54 Splaiul Independentei, 050094
Bucharest, Romania\\
$^{4}$ Department of Physical Electronics, Faculty of Engineering, Tel Aviv
University, Tel Aviv 69978, Israel}

\maketitle

\section{Introduction}

A well-known fact is that the formation of stable dissipative
solitons---most typically, in lasing media \cite{Rosa,lasers} and plasmonic
cavities \cite{plasmonics}---relies upon the simultaneous balance of
competing conservative and dissipative effects in the system, i.e.,
respectively, the diffraction and self-focusing nonlinearity, and linear and
nonlinear loss and gain \cite{DS}. The generic model describing media where
stable dissipative solitons emerge via this mechanism is based on the
complex Ginzburg-Landau (CGL) equations with the cubic-quintic (CQ)
combination of gain and loss terms, which act on top of the linear loss \cite%
{lasers}. In addition to modeling the laser-physics and plasmonic settings,
the CGL equations, including their CQ variety, serve as relevant models in
many other areas, well-known examples being Bose-Einstein condensates in
open systems (such as condensates of quasi-particles in solid-state media)
\cite{BEC}, reaction-diffusion systems \cite{MCCROSS}, and superconductivity
\cite{supercond}. Thus, the CGL equations constitute a class of universal
models for the description of nonlinear waves and pattern formation in
dissipative media \cite{AK}.

The CGL equation with the CQ nonlinearity was originally postulated by
Petviashvili and Sergeev \cite{Petv} as a model admitting stable localized
two-dimensional (2D) patterns. Subsequently, systems of this type were
derived or introduced phenomenologically in many physical settings, and a
great deal of 1D and 2D localized solutions, i.e., dissipative solitons,
have been studied in detail in them \cite{Boris}-\cite{Vladimir}.

A 2D model of laser cavities with an internal transverse grating, based on
the CQ-CGL equation supplemented by a spatially periodic (lattice)
potential, which represents the grating, was introduced in Ref. \cite%
{leblond1}. Note that the currently available laser-writing technology makes
it possible to fabricate permanent gratings in bulk media \cite{Jena}. In
addition, in photorefractive crystals virtual photonic lattices may be
induced by pairs of pump laser beams with the ordinary polarization, which
illuminate the sample in the directions of $x$ and $y$, while the probe beam
with the extraordinary polarization is launched along axis $z$ \cite%
{Moti-general}.  In fact, the laser cavity equipped with the grating may be
considered as a photonic crystal built in the active medium. 
Periodic potentials are also known in passive optical systems, driven by
external laser beams and operating in the temporal domain, unlike the
spatial-domain dynamics of the active systems. In such systems, effective
lattices may be induced by spatial modulation of the pump beam \cite%
{Firth,advances}. 

A notable fact reported in Ref. \cite{leblond1} is that localized vortices,
built as sets of four peaks pinned to the periodic potential, may be stable
without the presence of the diffusion term in the CGL laser model,which is
necessary for the stabilization of dissipative vortex solitons in uniform
media, see, e.g., Ref. \cite{Mihalache}, but is unphysical for waveguiding
models (the diffusion term is relevant in models describing light trapped in
a cavity, where the evolutional variable is time, rather than the
propagation distance \cite{Fedorov}). In subsequent works, stable
fundamental and vortical solitons in 2D \cite{trapping_potentials_2D} and 3D
\cite{trapping_potentials_3D} CGL models with trapping potentials were
studied in detail. Spatiotemporal dissipative solitons in the CQ-CGL\ model
of 3D laser cavities including the transverse grating were investigated too
\cite{trapping_potentials_3D}. Both fundamental and vortical solitons were
found in a numerical form as attractors in the latter model, and their
stability against strong random perturbations was tested by direct
simulations.

While the stability of various 2D localized patterns has been studied
thoroughly in the framework of the CQ-CGL equations with the transverse
lattice potential used as the stabilizing factor, a challenging problem is
mobility of such 2D dissipative solitons under the action of a kick applied
across the underlying lattice.

Actually, the action of the kick in this context implies the application of
a tilt to the beam. It should be noted that the CGL equation models
single-pass optical amplifiers, as well as laser cavities. In the latter
case the evolution is considered in the temporal domain, while in the former
the evolution variable is the propagation distance. In order to build the
required initial data in a laser cavity, the tilt should be applied very
quickly. This may be achieved by means of an optically induced grating or
mirrors, using some fast device. On the other hand, in the single-pass
amplifier the input is an incident beam, the tilt being a mere misalignment
between the beam and the amplifier's axis. Below, we present the analysis in
terms of the amplifier's setup, which is more straightforward for the
experimental implementation.

Thus, we assume that an external source produces a beam, which is shaped
into the fundamental-soliton mode of the amplifier by means of an adequate
setup. Then, the direction of this input beam is slightly tilted with
respect to the amplifier's axis, allowing the transverse part of the beam's
momentum to acts as the kick applied to the fundamental spatial soliton.

Furthermore, we demonstrate that the effective hopping motion of the kicked
soliton through cells of the periodic potential can be used for controlled
creation of various patterns filling these cells (or a part of them).

Thus, the main objective of this work is to study the mobility of the 2D
fundamental solitons, and scenarios of the pattern formation by kicked ones,
in the framework of the CQ-CGL models with the lattice potential. The model
is formulated in Section II, which also presents an analytical
approximation, that, using the concept of the Peierls-Nabarro (PN)\ barrier,
makes it possible to predict, with a reasonable accuracy, the minimum
(threshold) strength of the kick necessary for depinning of the quiescent
soliton trapped by the lattice. The main numerical results for the mobility
of the kicked soliton, and various scenarios of the pattern creation in the
wake of the soliton hopping between cells of the potential lattice, are
reported in Section III, while Section IV deals with collisions between a
freely moving soliton and a standing structure created and left by it, in
the case of periodic boundary conditions (which correspond to a pipe-shaped
amplifier, i.e., one in the form of a hollow cylinder). In particular,
elastic collisions provide for an example of a soliton Newton's cradle. The
paper is concluded by Section V.

\section{The model and analytical approximations}

\subsection{The Ginzburg-Landau equation}

Following Refs. \cite{leblond1} and \cite{trapping_potentials_2D}, the
scaled CQ-CGL equation for the evolution of the amplitude of the
electromagnetic field, $u(X,Y,Z)$, in two dimensions with transverse
coordinates $\mathbf{R}=(X,Y)$, along the propagation direction, $Z$, is
written as

\begin{equation}
\frac{\partial u}{\partial Z}=\left[ -\delta +iV(X,Y)+\frac{i}{2}\nabla
_{\perp }^{2}+(i+\epsilon )|u|^{2}-(i\nu +\mu )|u|^{4}\right] u,  \label{CGL}
\end{equation}%
where the paraxial diffraction is represented by $\nabla _{\perp
}^{2}=\partial ^{2}/\partial X^{2}+\partial ^{2}/\partial Y^{2}$ , real
coefficients $\delta $, $\epsilon $, and $\mu $ account for the linear loss,
cubic gain, and quintic loss, respectively, and coefficient $\nu >0$
accounts for the saturation of the Kerr nonlinearity. The transverse grating
is represented by the periodic potential,
\begin{equation}
V(X,Y)=-V_{0}\left[ \cos (2X)+\cos (2Y)\right] ,  \label{V}
\end{equation}%
of depth $2V_{0}$, with the period scaled to be $\pi $. Localized modes
produced by Eq. (\ref{CGL}), which physically correspond to light beams
self-trapped in the $\left( X,Y\right) $ plane, are characterized by the
total power,
\begin{equation}
P=\int \int |u(X,Y,Z)|^{2}dXdY.  \label{P}
\end{equation}

In the simulations, Eq. (\ref{CGL}) was solved by means of the standard
fourth-order Runge-Kutta scheme in the $Z$-direction, and five-point
finite-difference approximation for the transverse Laplacian. As specified
below, we used periodic boundary conditions. The integration domain
correspond to an $(X,Y)$ matrix of $256\times 256$ grid points covering the
area of $\left\vert X\right\vert ,\left\vert Y\right\vert \leq 22$. Generic
results for the mobility of fundamental dissipative solitons can be
adequately represented by fixing the following set of parameters:
\begin{equation}
\delta =0.4,\epsilon =1.85,\mu =1,\nu =0.1,V_{0}=1,  \label{set}
\end{equation}%
for which the quiescent fundamental soliton is stable. 

\subsection{The description of tilted beams}

In simulations of Eq. (\ref{CGL}), the kick (i.e., tilt) was applied to the
self-trapped beam, as usual, by multiplying the respective steady state, $%
u_{0}$, by $\exp \left( i\mathbf{k}_{0}\mathbf{R}\right) $, with the
vectorial strength of the kick defined as
\begin{equation}
\mathbf{k}_{0}=\left( k_{0}\cos \theta ,k_{0}\sin \theta \right) ,  \label{k}
\end{equation}%
where the square-lattice symmetry of potential (\ref{V}) makes it sufficient
to confine the orientation angle, $\theta $, to interval $0\leq \theta \leq
\pi /4$. In terms of the amplifier setup, a small angle $\varphi $ between
the carrier wave vector $\mathbf{K}_{0}=(K_{x},K_{y},K_{z})$ of the beam (in
physical units) and the $Z$-axis, gives rise to the transverse component $%
\mathbf{k}_{0}$ which corresponds to the kick. The paraxial approximation
implies that
\begin{equation}
\varphi \approx \sqrt{K_{x}^{2}+K_{y}^{2}}/K_{z}~\ll 1,  \label{parax}
\end{equation}
Before studying effects induced by the kick, it is relevant to explain the
corresponding physical setting in more detail, making it sure that the
tilted beams remain within the confines of the paraxial description.

Equation (\ref{CGL}) can be derived from the underlying wave equation by
means of the standard slowly varying envelope approximation (SVEA) \cite%
{Rosa}. For this purpose, electric field $E$ (in its scalar form) is split
into a slowly varying amplitude and the rapidly oscillating carrier either
as
\begin{equation}
E=\mathcal{A}(x,y,z-vt)e^{i\left( K_{x}x+K_{y}y+K_{z}z-\omega t\right) }+%
\mathrm{c.c.},  \label{dec1}
\end{equation}%
where $(x,y,z,t)$ are the coordinates and time in physical units, $v$ is the
group velocity, and $\mathrm{c.c.}$ stands the complex conjugate, or,
alternatively, as
\begin{equation}
E=A(x,y,z-vt)e^{i\left( K_{z}z-\omega t\right) }+\mathrm{c.c.}  \label{dec2}
\end{equation}%
The difference is that that carrier wave is oblique in Eq. (\ref{dec1}),
while in Eq. (\ref{dec2}) it is always defined as the straight one.
Obviously, $A(x,y,z-vt)=\mathcal{A}(x,y,z-vt)e^{i\left( K_{x}x+K_{y}y\right)
}$, and the two forms are fully equivalent within the framework of the SVEA
if the oscillations due to the term $e^{i\left( K_{x}x+K_{y}y\right) }$ are
not essentially faster than the transverse variations of the beam described
by amplitude $\mathcal{A}(x,y,z-vt)$. Thus, the SVEA can be fixed in the
form of Eq. (\ref{dec2}), with $K_{z}=2\pi n/\lambda $, where $n$ the linear
index of the medium.

Further, to proceed to normalized equation (\ref{CGL}), we set $\left\{
X,Y\right\} \equiv w^{-1}\left\{ x,y\right\} $, where $w$ is a
characteristic scale of the beam's width---say, $w\sim 10$ $\mathrm{\mu }$m
for narrow beams, if the underlying wavelength (in vacuum) is $\lambda \sim
1 $ $\mathrm{\mu }$m. Accordingly, the period of grating (\ref{V}) is $\pi w$
in the physical units. Note that $w\sim 10$ $\mathrm{\mu }$m corresponds to
the period $\pi w\sim 30$ $\mathrm{\mu }$m, which is relevant estimate
(gratings with the period on this order of magnitude can be readily
manufactured). Then, the scaled propagation distance is $Z=z/(K_{z}w)\equiv
\left( \lambda /2\pi nw\right) z$, and the scale wave amplitude is%
\[
u=\sqrt{\frac{1}{2}\mathrm{Re}\left( \chi ^{(3)}\right) }\frac{\omega w}{c}%
A,
\]%
where $\chi ^{(3)}$ is the third-order nonlinear susceptibility. Further,
the respective rescaling of the wave vector components is given by $\left\{
k_{x},k_{y}\right\} =w\left\{ K_{x},K_{y}\right\} $. Thus, the deviation
angle can be estimated as $\tan \varphi =\sqrt{K_{x}^{2}+K_{y}^{2}}%
/K_{z}\equiv \left( \lambda /2\pi nw\right) \sqrt{k_{x}^{2}+k_{y}^{2}}$,
and, for generic tilted modes, with $\sqrt{k_{x}^{2}+k_{y}^{2}}\sim 1$ in
the scaled notation, condition (\ref{parax}) for the validity of the
paraxial approximation amounts to $\lambda \ll 2\pi nw$, which is nothing
but the standard paraxial assumption.

With the above-mentioned typical values, $\lambda \sim 1$ $\mathrm{\mu }$m
and $w\sim 10$ $\mathrm{\mu }$m, along with $n\approx 1.5$, the above
estimate yields $\varphi \sim 0.01$ (in radians). Then, the minimum
propagation distance relevant to the experiment, $z\sim 1$ cm, corresponds
to the transverse deviation of the tilted beam $\Delta x\sim 100$ $\mathrm{%
\mu }$m, which can be easily detected and employed in applications.

\subsection{Analytical estimates}

The first characteristic of kink-induced effects is the threshold value, $%
\left( k_{0}\right) _{\mathrm{thr}}$, such that the soliton remains pinned
at $k_{0}<\left( k_{0}\right) _{\mathrm{thr}}$ and escapes at $k_{0}>\left(
k_{0}\right) _{\mathrm{thr}}$.To develop an analytical approximation which
aims to predict the threshold, one can, at the lowest order, drop the loss
and gain terms, as well as the lattice potential, in Eq. (\ref{CGL}). The
corresponding 2D nonlinear Schr\"{o}dinger equation gives rise to the
commonly known family of Townes soliton, which share a single value of the
total power, $P_{\mathrm{T}}\approx 5.85$ \cite{Berge'}. The family may be
approximated by the isotropic Gaussian ansatz with arbitrary amplitude $A$,%
\begin{equation}
u\left( Z,R\equiv \sqrt{X^{2}+Y^{2}}\right) =A\exp \left( ibZ-\frac{\pi A^{2}%
}{2P_{\mathrm{T}}}R^{2}\right) ,  \label{ans}
\end{equation}%
and propagation constant $b=A^{2}/4$ \cite{Anderson}. Then, taking into
regard the loss and gain terms as perturbations, one can predict the
equilibrium value of the amplitude from the power-balance equation,%
\begin{equation}
\delta \cdot P+\mu \int \int \left\vert u\left( X,Y\right) \right\vert
^{6}dXdY=\epsilon \int \int \left\vert u\left( X,Y\right) \right\vert
^{4}dXdY.  \label{balance}
\end{equation}%
The substitution of approximation (\ref{ans}) into Eq. (\ref{balance}) leads
to a quadratic equation for $A^{2}$, with roots%
\begin{equation}
A^{2}=\frac{3\epsilon \pm \sqrt{3\left( 3\epsilon ^{2}-16\mu \delta \right) }%
}{4\mu },  \label{root}
\end{equation}%
the larger one corresponding to a stable dissipative soliton (cf. a similar
analysis for the CQ-CGL model in 1D \cite{PhysicaD}).

Proceeding to the kicked soliton, the threshold magnitude of the kick for
the depinning, $\left( k_{0}\right) _{\mathrm{thr}}$, can be estimated from
the comparison of the PN potential barrier, $U_{\mathrm{PN}}$, and the
kinetic energy of the kicked soliton, $E_{\mathrm{kin}}$. The Galilean
invariance of Eq. (\ref{CGL}) \cite{HS} (in the absence of the lattice
potential) implies that kick (\ref{k}) gives rise to ``velocity" $\mathbf{k}%
_{0}$, so that the solution will become a function of $\mathbf{R}-\mathbf{k}%
_{0}Z\equiv \mathbf{R-}\mathbf{\Upsilon }$, instead of $\mathbf{R}$, the
corresponding kinetic energy being%
\begin{equation}
E_{\mathrm{kin}}=(1/2)Pk_{0}^{2}  \label{kin}
\end{equation}%
($P$ plays the role of the effective mass of the soliton). Further, assuming
that the kicked soliton moves in the direction of $\theta $ [see Eq. (\ref{k}%
)], the effective energy of the interaction of the soliton, taken as per
approximation (\ref{ans}), with lattice potential (\ref{V}), treated as
another perturbation, is%
\begin{eqnarray}
E_{\mathrm{pot}}(\Upsilon ) &=&-V_{0}\int \int \left[ \cos (2X)+\cos (2Y)%
\right] \left\vert u\left( X-\Upsilon \cos \theta ,Y-\Upsilon \sin \theta
\right) \right\vert ^{2}dXdY  \nonumber \\
&=&-V_{0}P\exp \left( -\frac{P_{\mathrm{T}}}{\pi A^{2}}\right) \left[ \cos
\left( 2\Upsilon \cos \theta \right) +\cos \left( 2\Upsilon \sin \theta
\right) \right] ,  \label{pot}
\end{eqnarray}%
where $\Upsilon $ is the shift of the soliton from $X=Y=0$ in the direction
of $\theta $. As follows from this expression, the PN\ barrier, i.e., the
difference between the largest and smallest values of the potential energy,
is estimated as%
\begin{equation}
E_{\mathrm{PN}}=V_{0}P\exp \left( -\frac{P_{\mathrm{T}}}{\pi A^{2}}\right)
\Delta (\theta ),  \label{PN}
\end{equation}%
where $\Delta (\theta )$ is the difference between the maximum and minimum
of function $\cos \left( 2\Upsilon \cos \theta \right) +\cos \left(
2\Upsilon \sin \theta \right) $. Obviously, $\Delta (0)=2$ and $\Delta
\left( \pi /4\right) =4$. For intermediate values of $\theta $, it may be
approximated by the difference of the values of the function between points $%
\Upsilon =0$ and $\Upsilon =\pi /\left( 2\cos \theta \right) $, i.e.,%
\begin{equation}
\Delta (\theta )\approx 3-\cos \left( \pi \tan \theta \right) .
\label{Delta}
\end{equation}%
Finally, the threshold value of the kick is determined by the depinning
condition, $E_{\mathrm{kin}}=E_{\mathrm{PN}}$, i.e.,%
\begin{equation}
\left( k_{0}\right) _{\mathrm{thr}}=\sqrt{2V_{0}\left[ 3-\cos \left( \pi
\tan \theta \right) \right] }\exp \left( -\frac{P_{\mathrm{T}}}{2\pi A^{2}}%
\right) .  \label{analyt}
\end{equation}%
This prediction is compared with numerical results below.

\section{Numerical results: The mobility and pattern formation}

\begin{figure}[th]
\centering%
\subfigure[Amplitude
]{\label{abssolfondinit}\includegraphics[height=3.5cm]{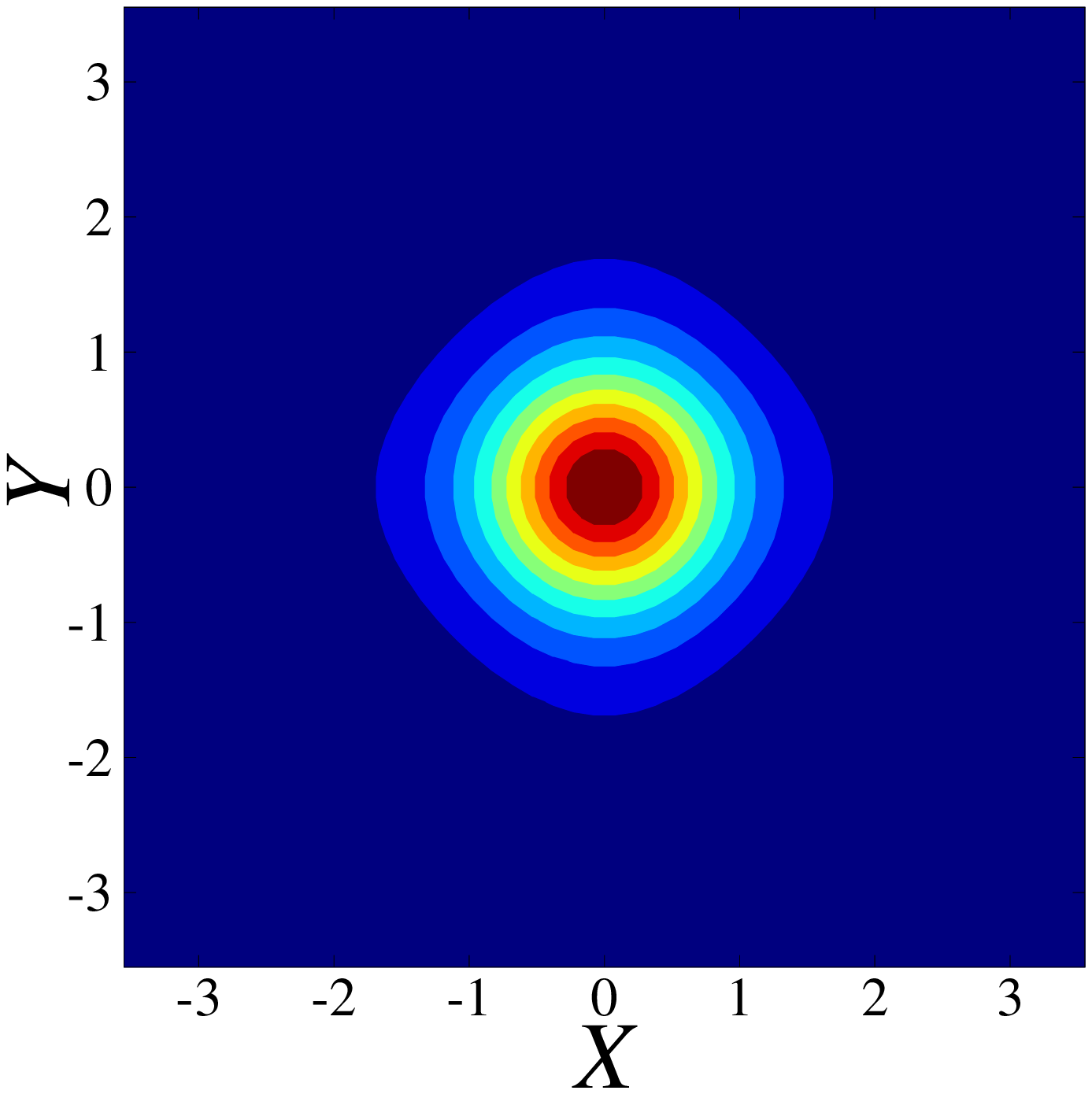}}
\subfigure[Profile
]{\label{profilsolfondinit}\includegraphics[height=3.5cm]{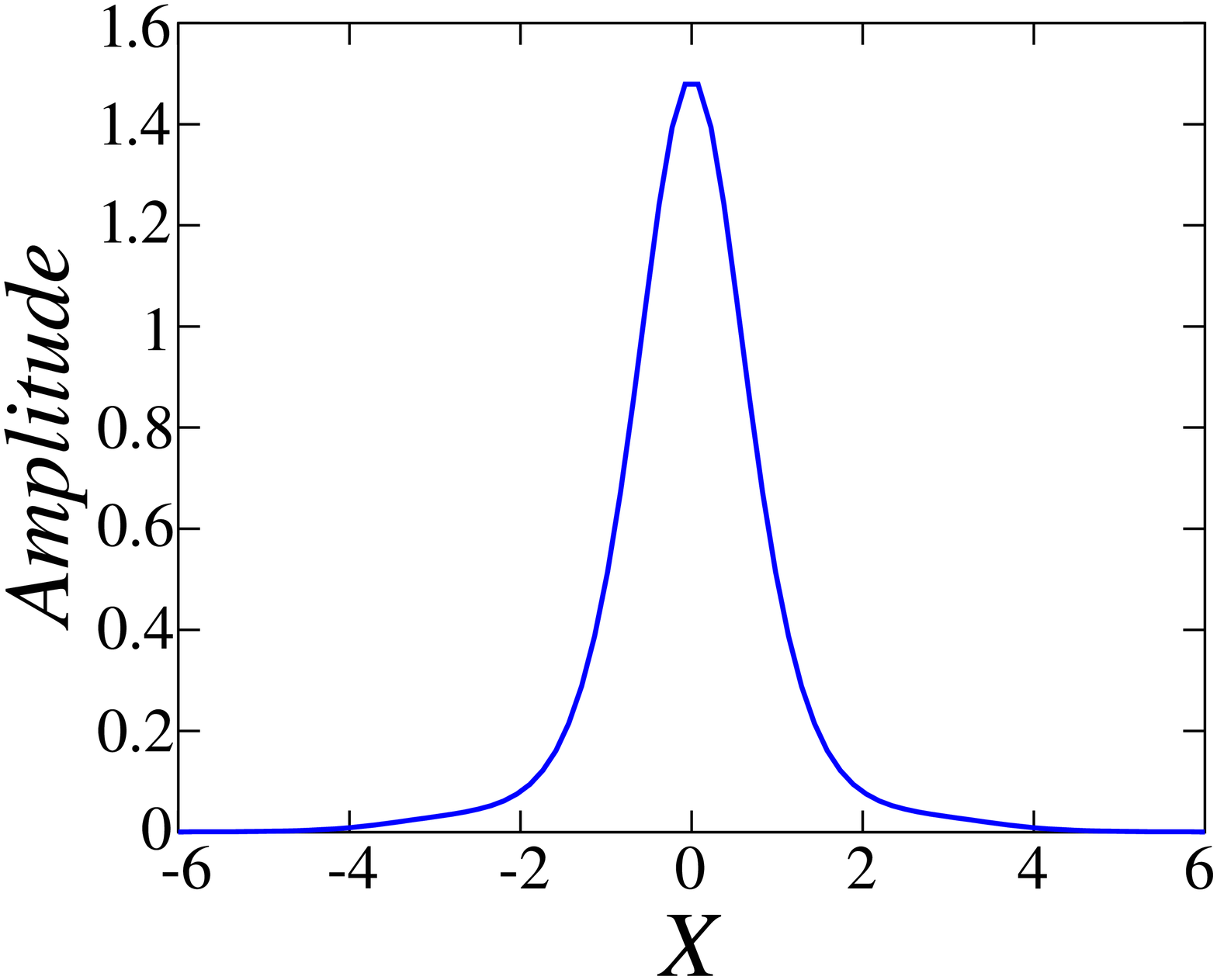}}
\caption{(Color online) The stable fundamental soliton: (a) the contour plot
of the local amplitude, $|u\left( X,Y\right) |$; (b) the cross-section
profile of $|u(X)|$ at $Y=0$. The soliton emphasizes the potential : $V(X,Y)=-%
\left[\cos\left(2X\right)+\sin\left(2X\right)\right]$ }
\label{solfondinit}
\end{figure}

The stable fundamental soliton constructed in the model based on Eqs. (\ref%
{CGL}) and (\ref{V}) at parameter values (\ref{set}) is shown in Fig. \ref%
{solfondinit}. For these parameters, analytical prediction (\ref{root})
yields the amplitude of the stable soliton $A\approx $ $\allowbreak 1.496\,$%
, which is quite close to the amplitude of the numerically found solution in
Fig. \ref{solfondinit}: $A\approx $ $\allowbreak 1.479\,$, which implies
that the isotropic Gaussian (\ref{ans}) is quite appropriate as the ansatz
for the description of static properties of the solitons.

\subsection{The formation of arrayed soliton patterns}

\begin{figure}[th]
\centering\subfigure[]{\label{absusolfon161}%
\includegraphics[width=4cm]{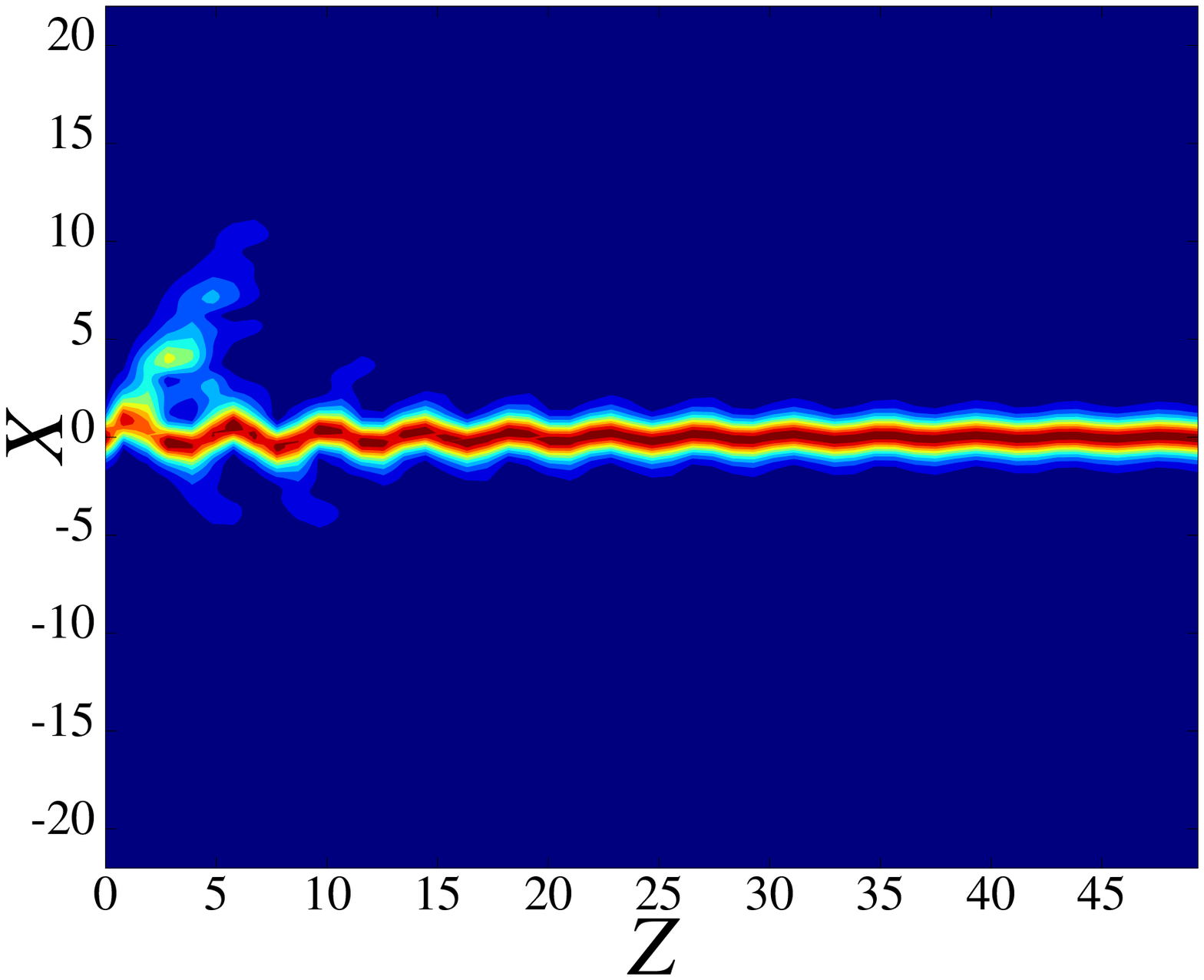}}
\subfigure[]{\label{energsolfon161}
\includegraphics[width=4cm]{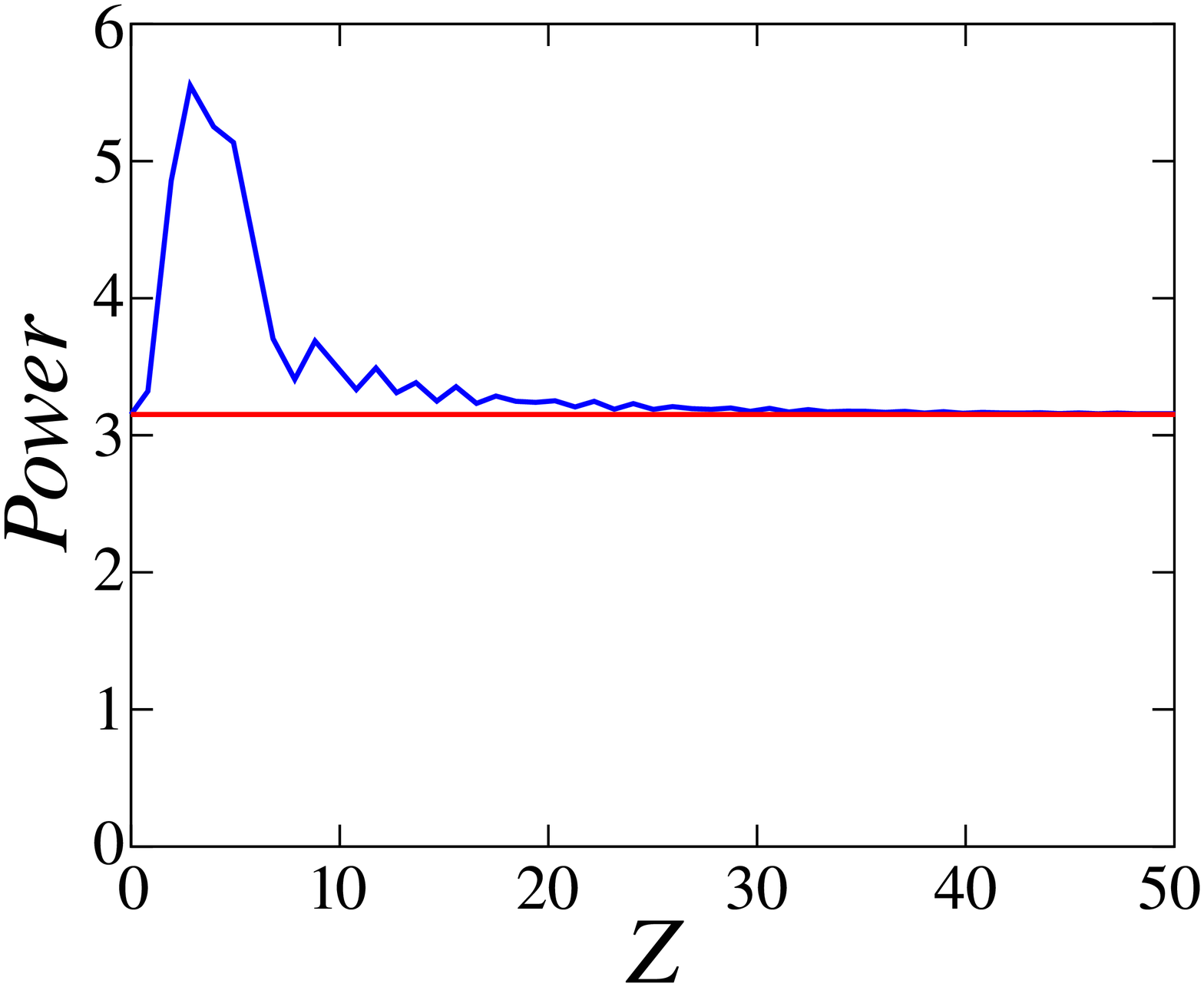}}
\caption{(Color online) (a): The evolution of the local amplitude, $|u(X,Z)|$%
, in the cross section, $Y=0$, of the fundamental soliton kicked with the
under-threshold strength, $k_{0}=1.61$ at $\protect\theta =0$. (b) The
evolution of the total power in this case. The horizontal line designates
the power of the quiescent fundamental soliton.}
\label{solfonn161}
\end{figure}

First we consider the solitons kicked with $\theta =0$, i.e., along bonds of
the lattice, see Eq. (\ref{k}). Below the threshold value of the kick's
strength, whose numerically found value is
\begin{equation}
\left( k_{0}\right) _{\mathrm{thr}}(\theta =0)\approx 1.6865,  \label{thr}
\end{equation}%
the kicked soliton exhibits damped oscillations, remaining trapped near a
local minimum of the lattice potential, as shown in Fig. \ref{solfonn161}.
Originally (at $0<z<8$ in Fig. \ref{solfonn161}), the total power (\ref{P})
increases, and then it drops to the initial value, $P\approx 3.2$. As a
result of the kick, a portion of the wave field passes the potential barrier
and penetrates into the adjacent lattice cell, but, at $k_{0}<\left(
k_{0}\right) _{\mathrm{thr}}$, the power carried by the penetrating field is
not sufficient to create a new soliton, and is eventually absorbed by the
medium.

On the other hand, analytical prediction (\ref{analyt}) yields $\left(
k_{0}\right) _{\mathrm{thr}}(\theta =0)\approx 1.32$. A relative discrepancy
$\simeq 20\%$ with numerical value (\ref{thr}) is explained by the fact
that, near the depinning threshold, the moving soliton suffers appreciable
deformation, while the analytical approach assumed the fixed shape (\ref{ans}%
), and it did not take into account energy losses (the latter factor makes
the actual threshold somewhat higher). In other cases considered below, see
Eqs. (\ref{pi/4}) and (\ref{pi/8}), the analytical predictions for $\left(
k_{0}\right) _{\mathrm{thr}}$ is also $\simeq 20\%$ smaller than their
numerically found counterparts

If the kick is sufficiently strong, $k_{0}>\left( k_{0}\right) _{\mathrm{thr}%
}$, the portion of the wave field passing the potential barrier has enough
power to create a new dissipative soliton in the adjacent cell. The emerging
secondary soliton may either stay in its cell, or keep moving through the
lattice.

\begin{figure}[th]
\begin{center}
\subfigure[$Z=2.65$]{\label{abssolfondk017theta0piZ5}%
\includegraphics[width=4cm]{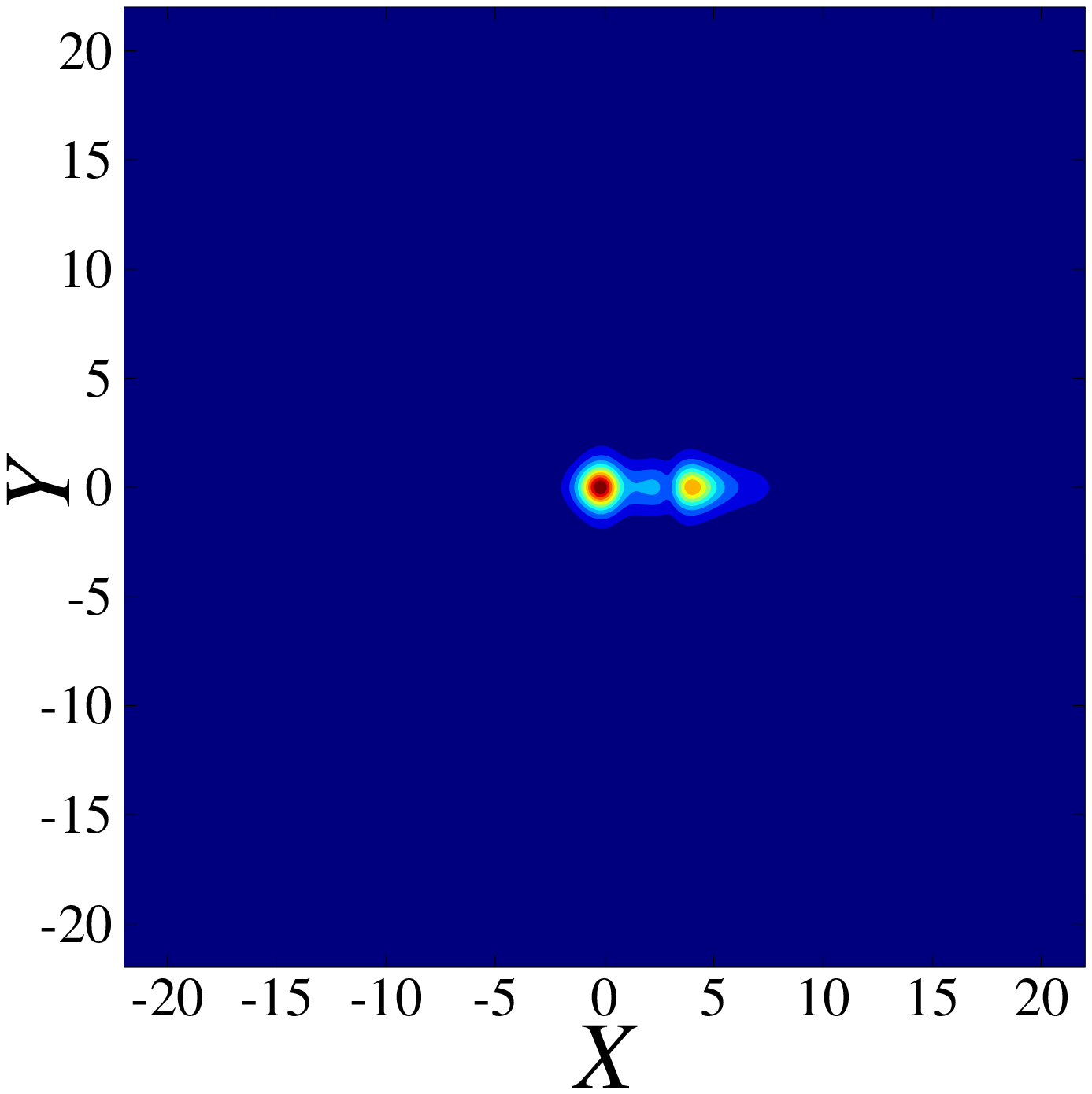}} %
\subfigure[$Z=4.70$]{\label{abssolfondk017theta0piZ15}%
\includegraphics[width=4cm]{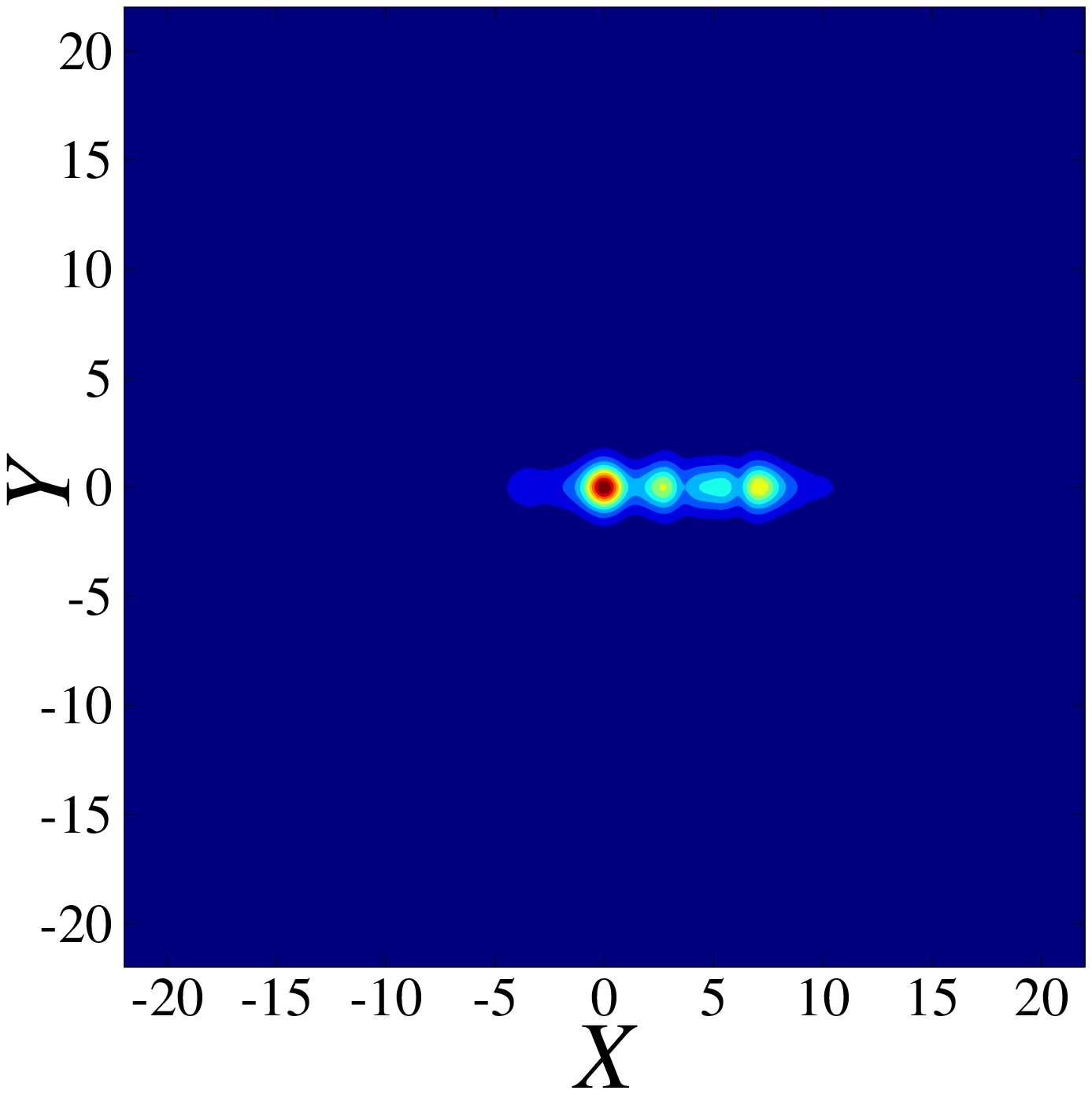}}\vfill
\par
\subfigure[$Z=8.51$]{\label{abssolfondk017theta0piZ25}%
\includegraphics[width=4cm]{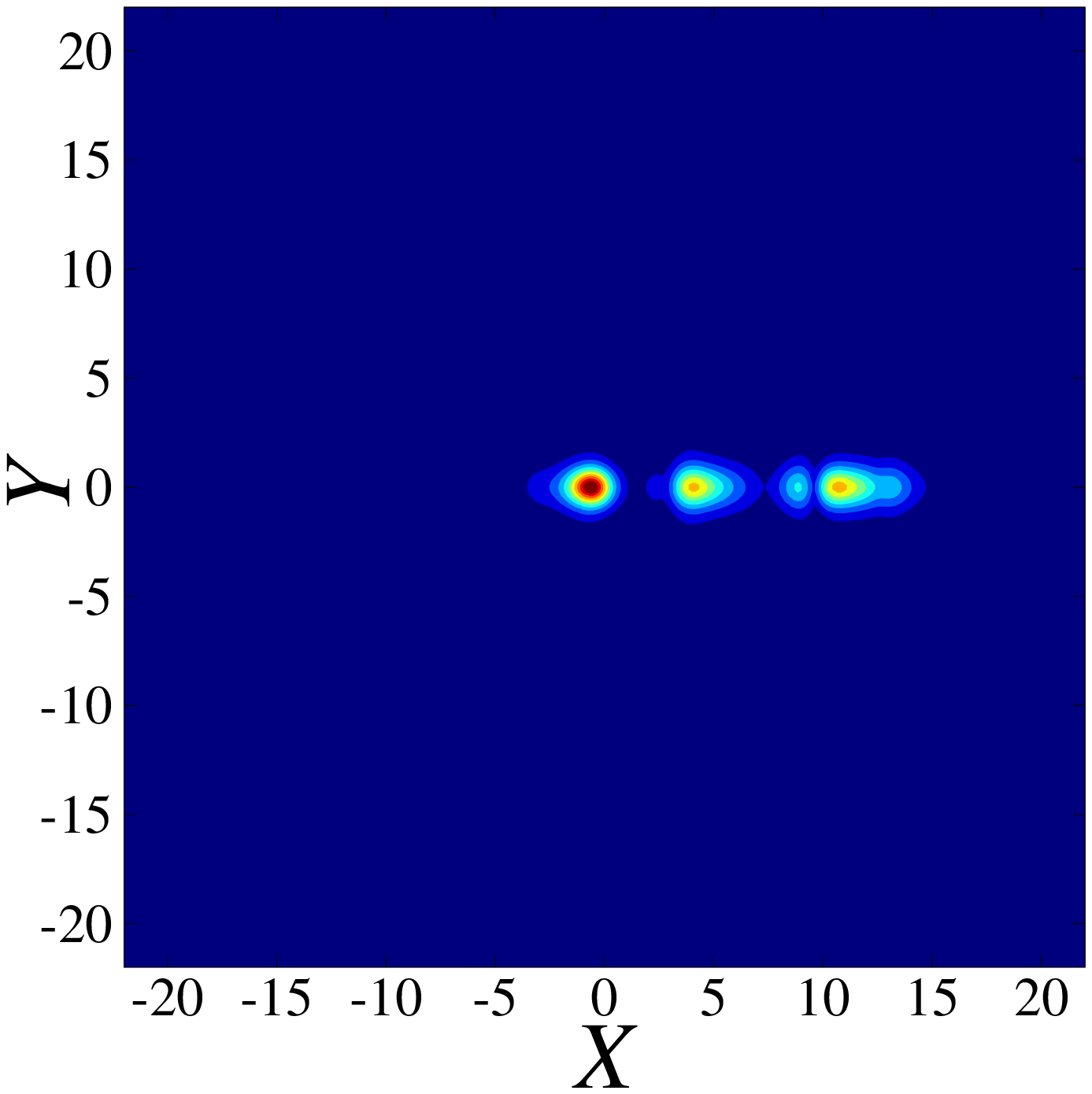}} %
\subfigure[$Z=14.00$]{\label{abssolfondk017theta0piZ35}%
\includegraphics[width=4cm]{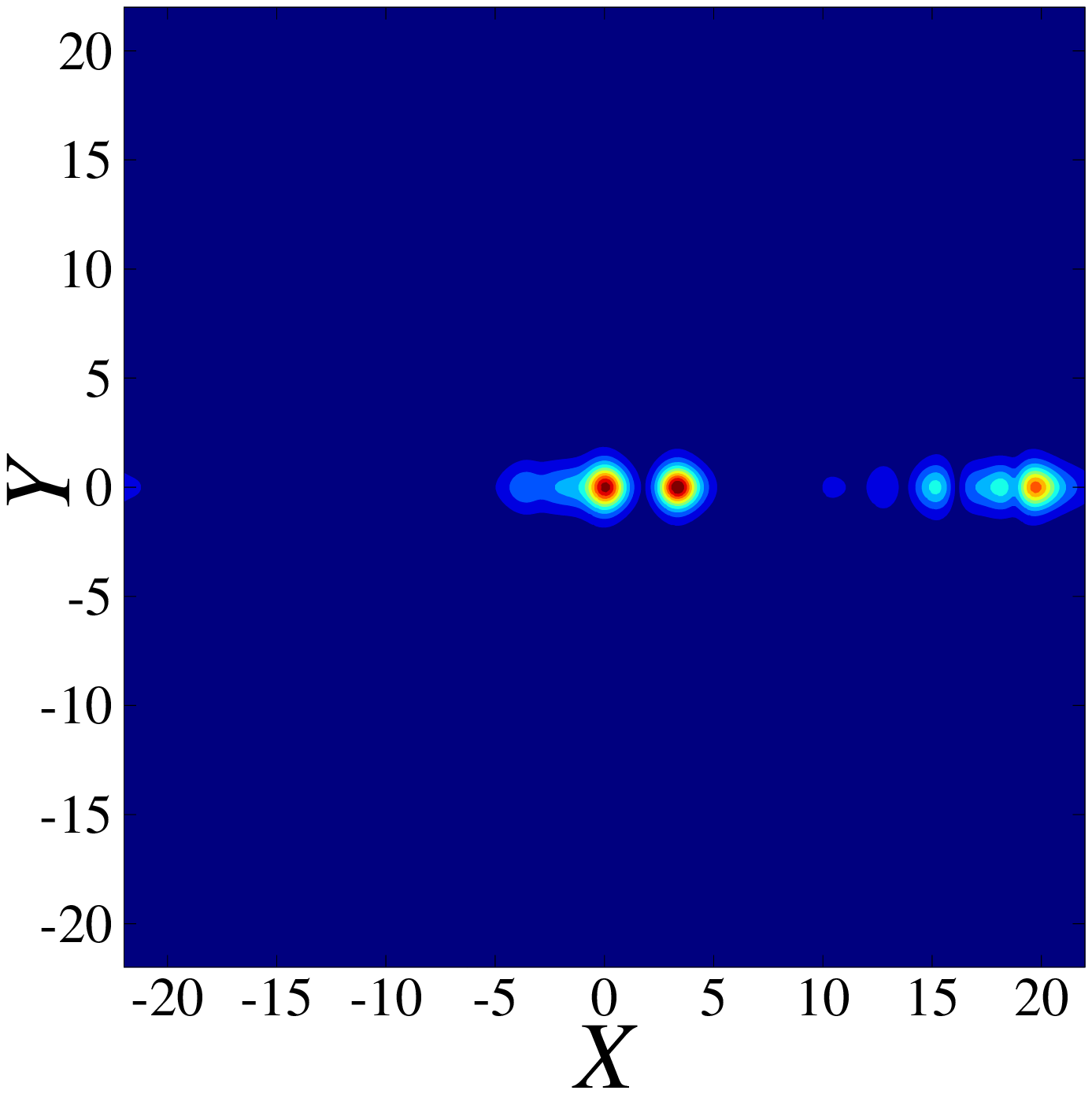}}
\end{center}
\caption{(Color online) The evolution of the local field amplitude, $%
|u(X,Y)| $, corresponding to the kicked soliton, for $k_{0}=1.6878$ and $%
\protect\theta =0$. The field distributions are displayed at different
values of propagation distance $Z$.}
\label{fond3}
\end{figure}

Figure \ref{fond3} demonstrates the creation of two new solitons at $%
k_{0}=1.6878$, which slightly exceeds the threshold value (\ref{thr}). This
figure represents a generic dynamical scenario, that can be summarized as
follows:

\begin{itemize}
\item[$\bullet $] the initial soliton (or a part of it) passes the potential
barrier and gets into the adjacent (second) cell;

\item[$\bullet $] it then stays for some time in that cell;

\item[$\bullet $] if the initial kick is not strong enough, the secondary
soliton permanently stays at this location;

\item[$\bullet $] if the kick is harder, the soliton again passes the
potential barrier, getting into the third cell, and may continue to move
through the grating;

\item[$\bullet $] a portion of the wave field of the secondary soliton stays
in the second cell and grows into a full soliton in this cell;

\item[$\bullet $] if the kick is not sufficient to continue the filling of
farther cells, oscillations of all the persistent solitons relax.
\end{itemize}

The present case is further illustrated in Fig. \ref%
{energsolfondk017theta0pi} by the plot for the evolution of the total power,
which shows that $P$ attains the first maximum at $Z=15.89$, and then
oscillates. Every minimum correspond to the collision between the two
solitons. For the sake of comparison, four horizontal lines in the figure
mark the powers corresponding to the single stable soliton ($P_{\mathrm{sol}%
}\approx 3.15$), multiplied, respectively, by $\ 1$, $2$, $3$, or $4$.

\begin{figure}[th]
\centering
\includegraphics[width=5cm]{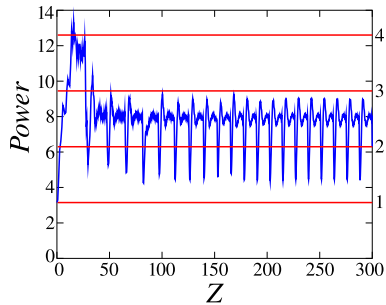}
\caption{(Color online) The evolution of the total power of the pattern in
the case displayed in Fig. \protect\ref{fond3}, with $k_{0}=1.6878$ and $%
\protect\theta =0$. The horizontal lines show the powers for the sets of 1,
2, 3 and 4 quiescent fundamental solitons.}
\label{energsolfondk017theta0pi}
\end{figure}

\begin{figure}[th]
\subfigure[$Z=2.65$]{\label{abssolfondk018theta0piZ5}%
\includegraphics[width=4cm]{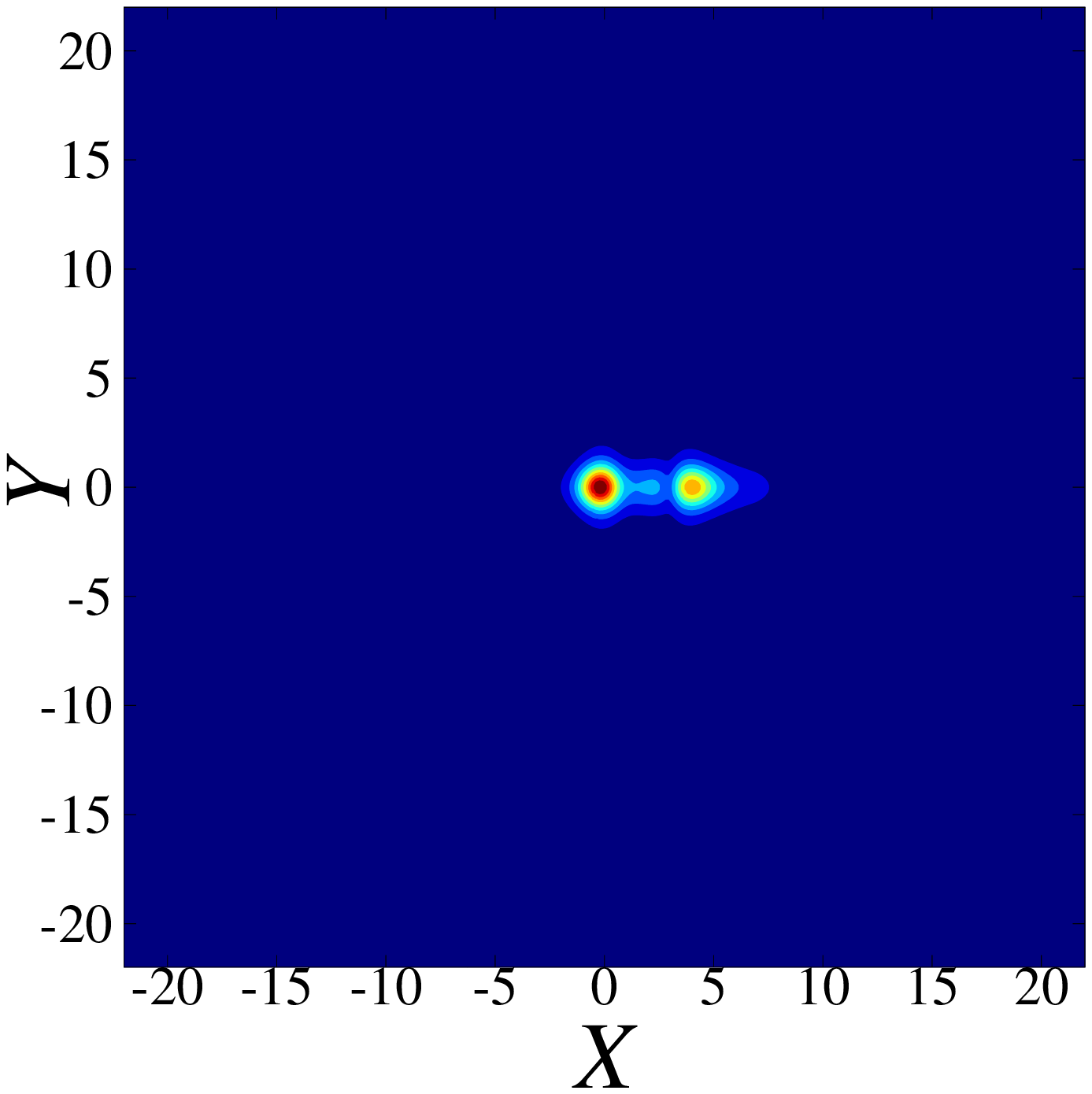}} %
\subfigure[$Z=8.51$]{\label{abssolfondk018theta0piZ15}%
\includegraphics[width=4cm]{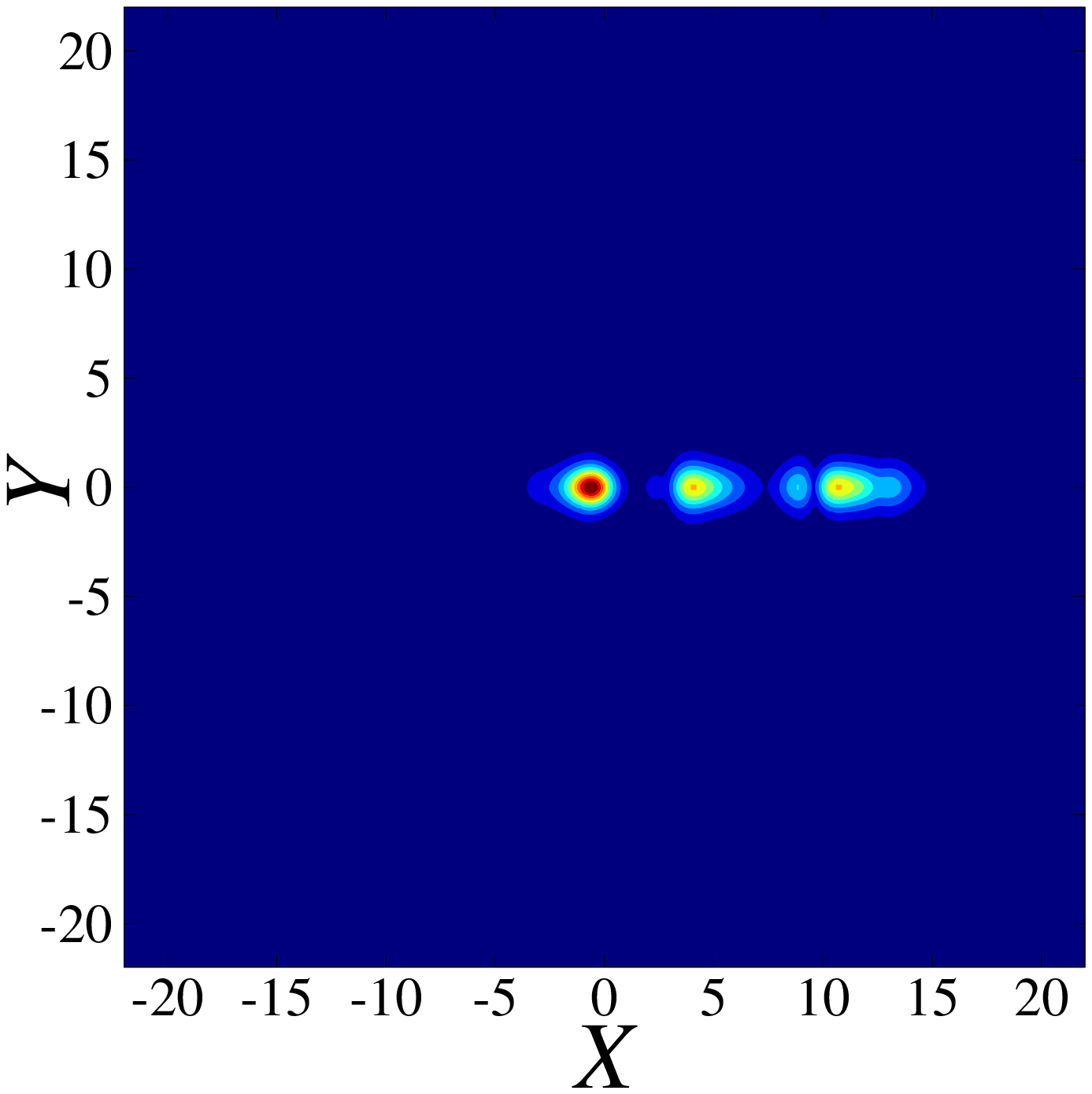}}\vfill
\subfigure[$Z=13.28$]{\label{abssolfondk018theta0piZ20}%
\includegraphics[width=4cm]{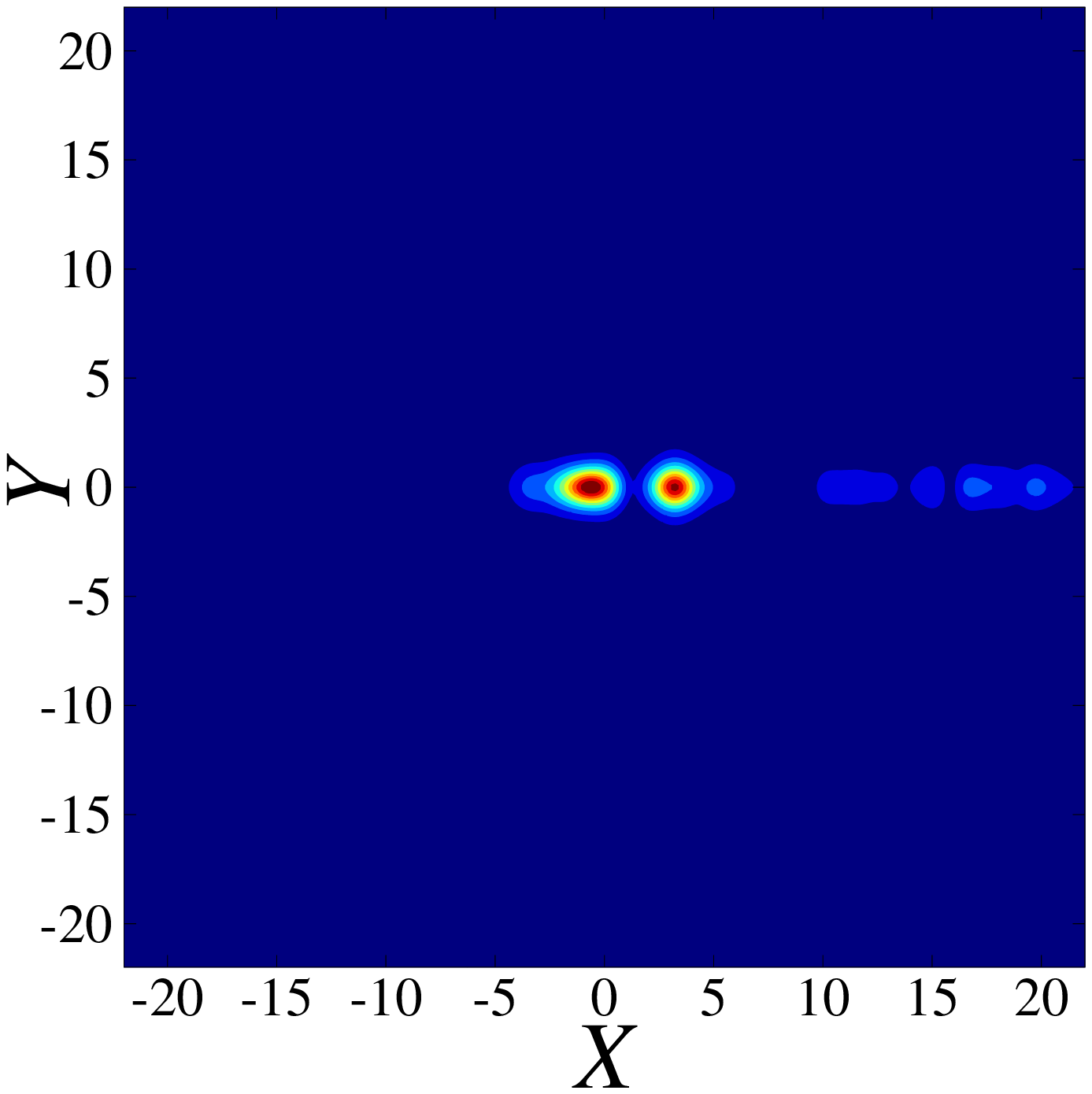}} %
\subfigure[$Z=21.43$]{\label{abssolfondk018theta0piZ25}%
\includegraphics[width=4cm]{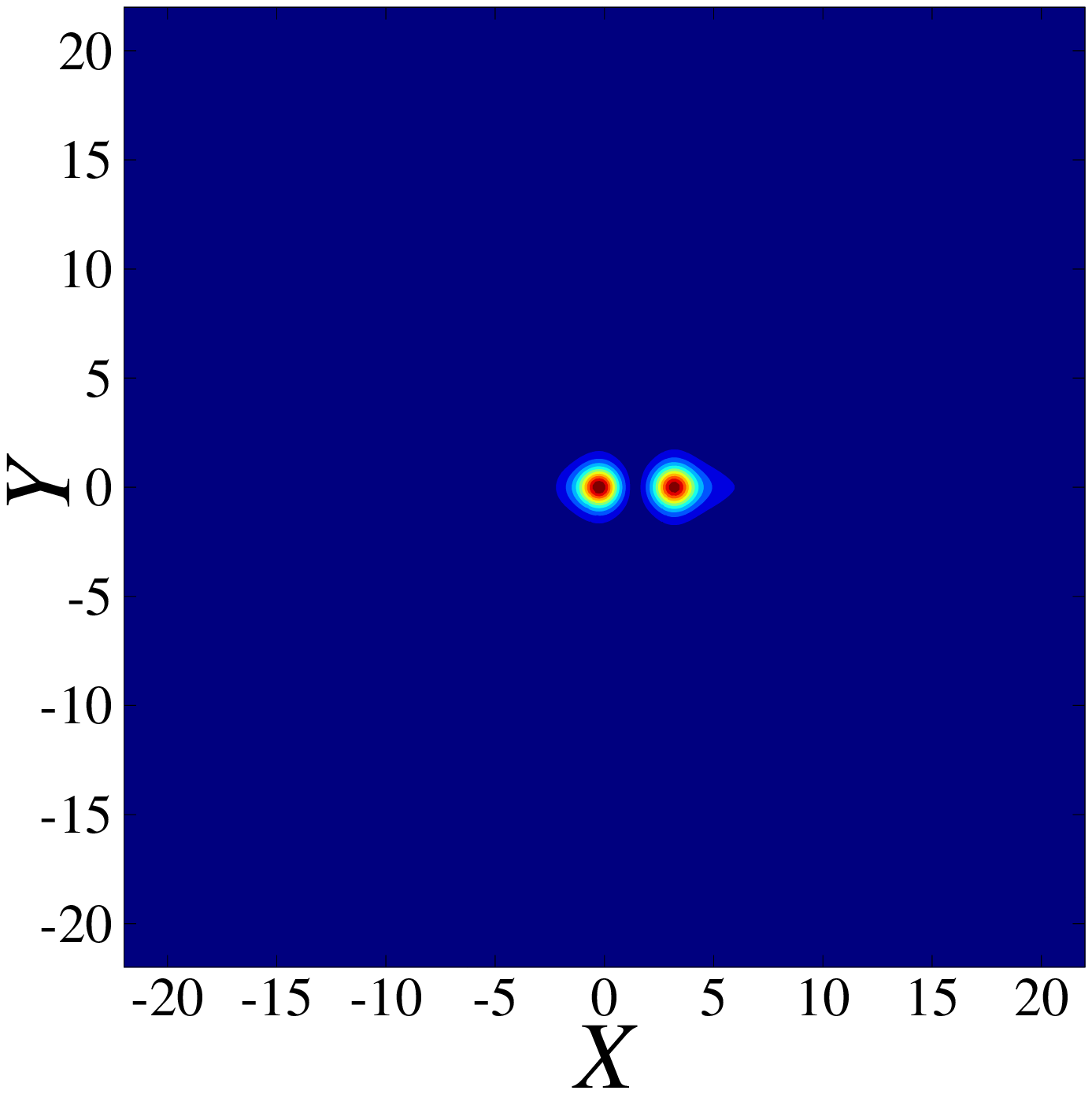}}
\caption{(Color online) The same as in Fig. \protect\ref{fond3}, but for $%
k_{0}=1.6872$ and $\protect\theta =0$.}
\label{fond4}
\end{figure}

It has been found that solitons can duplicate several times, thus forming
extended patterns in the form of soliton arrays. The increase of the kick's
magnitude leads to the decrease of the number of the solitons forming this
pattern, as the soliton moves faster and does not spend enough time in each
cell to create a new soliton trapped in it. In particular, Fig. \ref{fond4}
demonstrates that only one additional soliton is generated at $k_{0}=1.6872$%
, both solitons remaining pinned (note that this value is smaller than $%
k_{0}=1.6878$ appertaining to Fig. \ref{fond3}). Further, at $k_{0}=2.082$
the soliton performs unhindered motion, without leaving any stable pattern
in its wake (not shown here in detail).

\begin{figure}[th]
\begin{center}
\subfigure[$Z=2.63$]{\label{abssolfondk016theta0piZ5}%
\includegraphics[width=4cm]{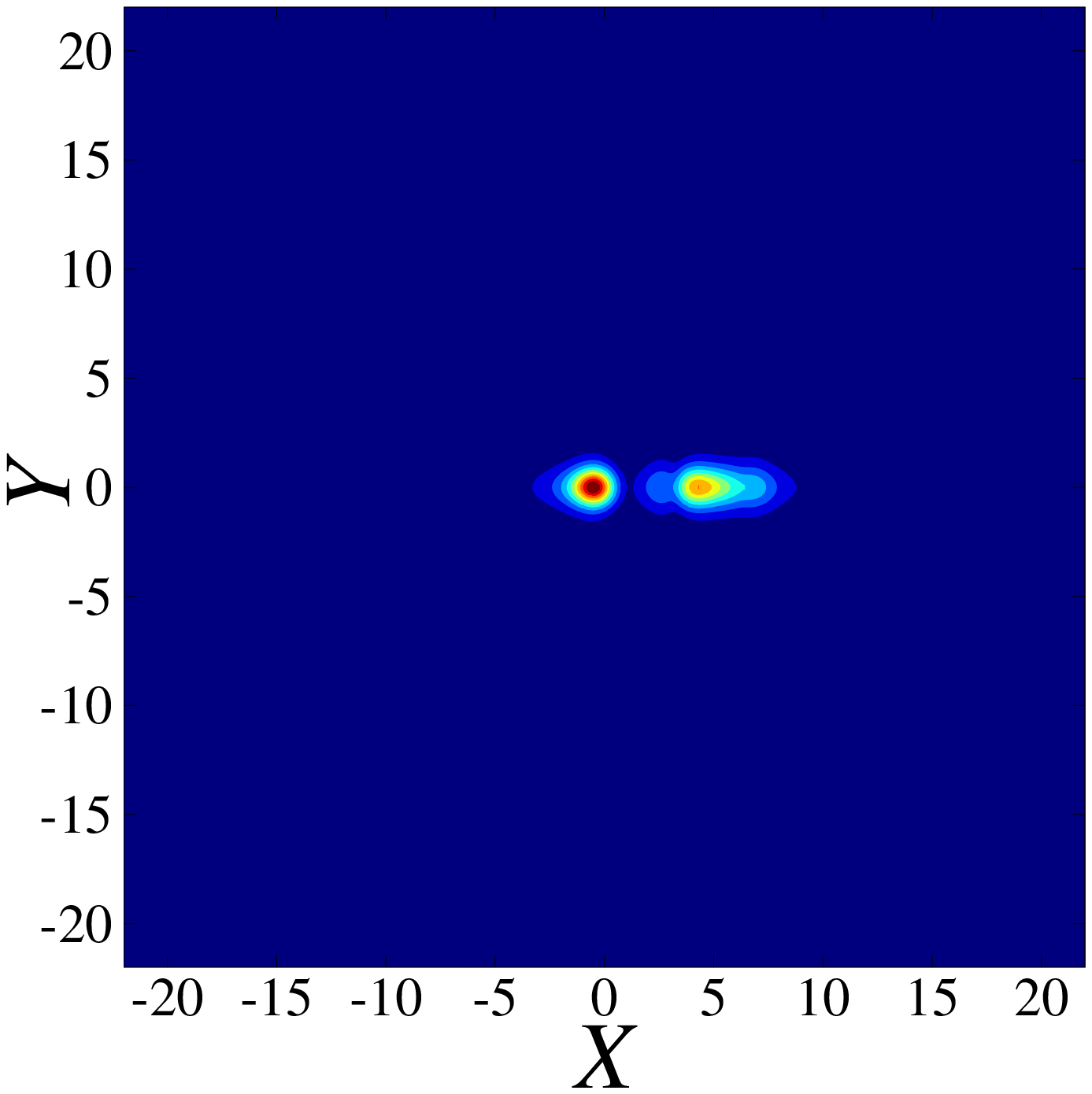}} %
\subfigure[$Z=4.68$]{\label{abssolfondk016theta0piZ10}%
\includegraphics[width=4cm]{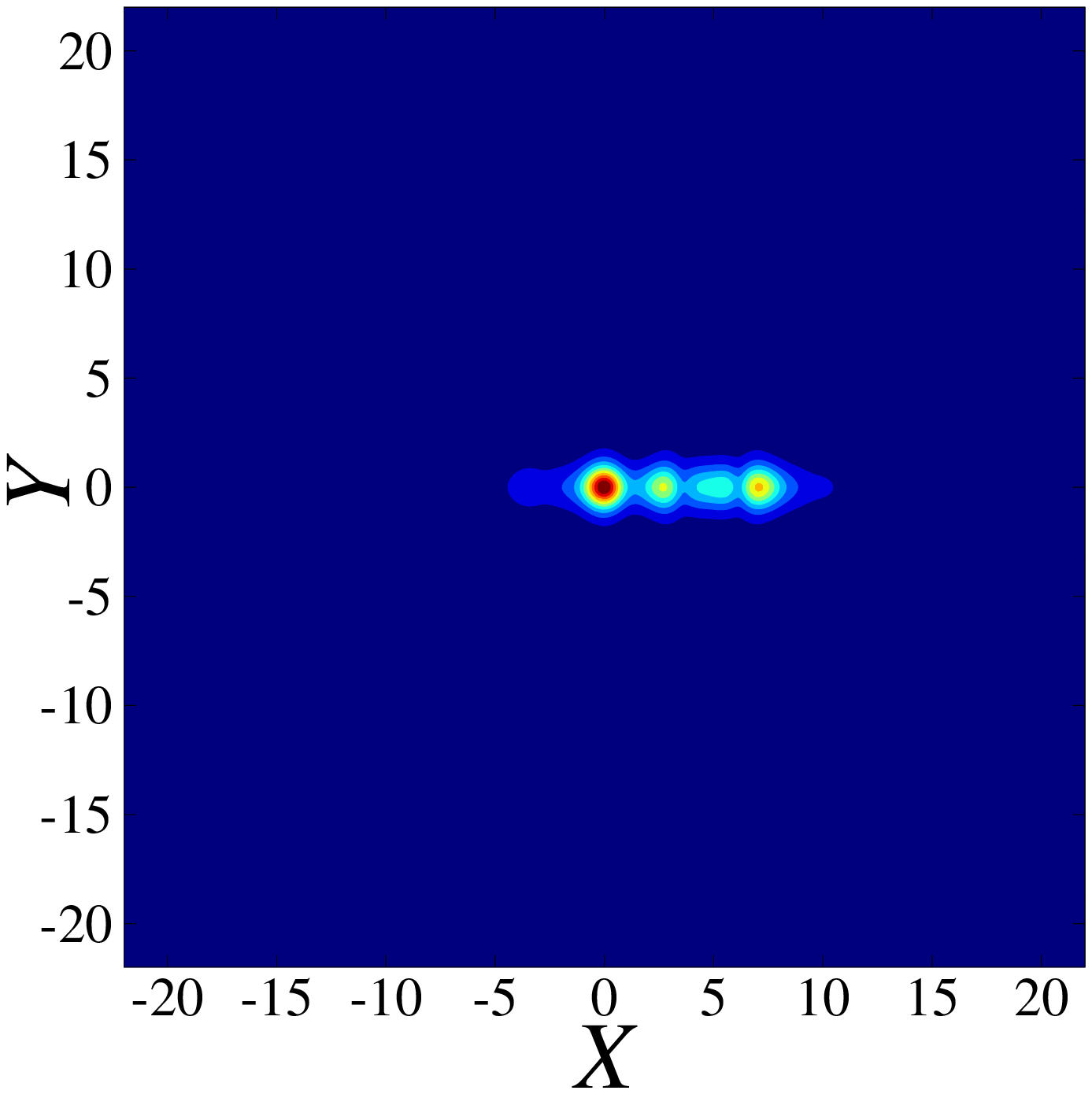}}\vfill
\subfigure[$Z=8.48$]{\label{abssolfondk016theta0piZ15}%
\includegraphics[width=4cm]{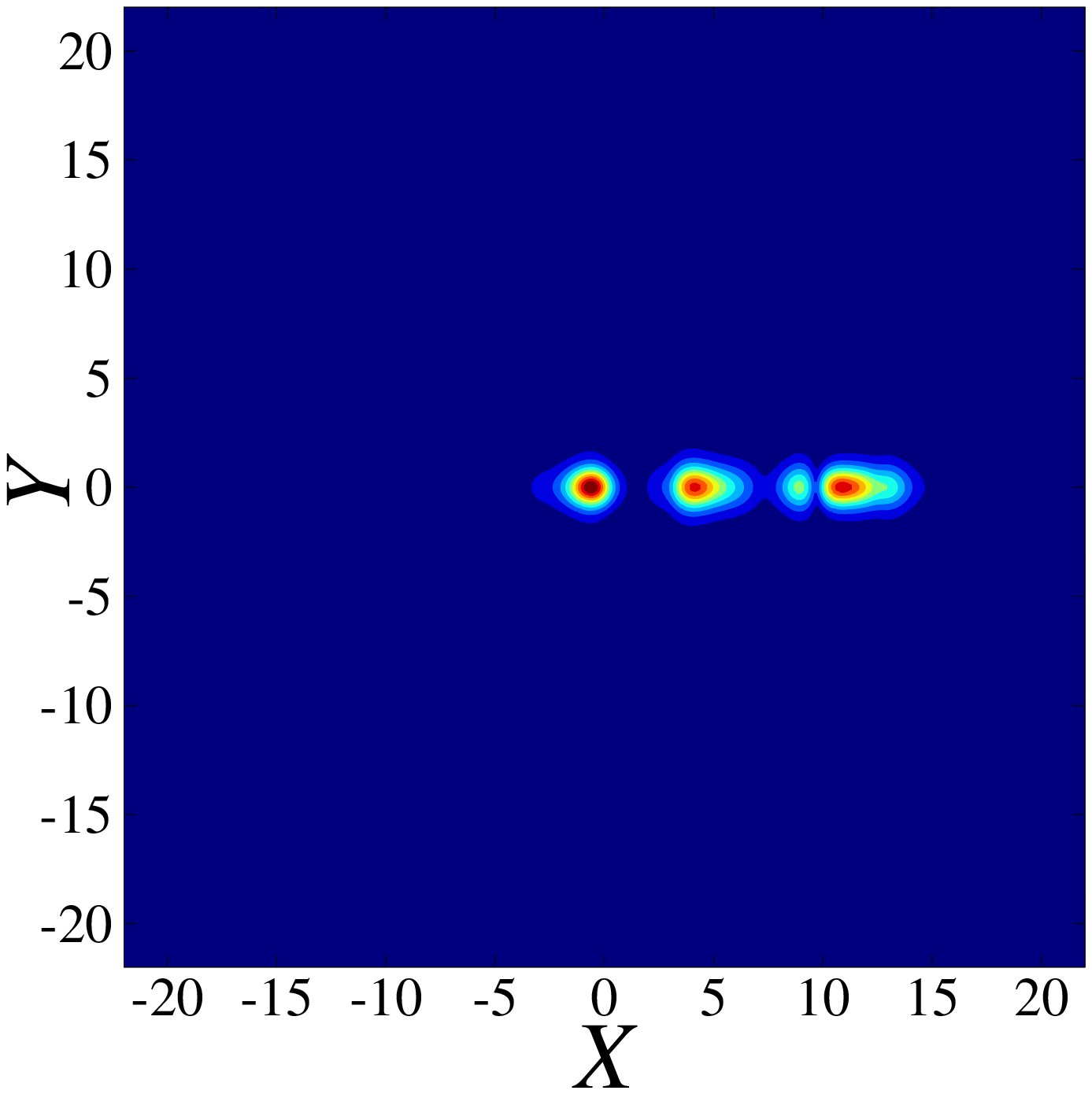}} %
\subfigure[$Z=12.05$]{\label{abssolfondk016theta0piZ25}%
\includegraphics[width=4cm]{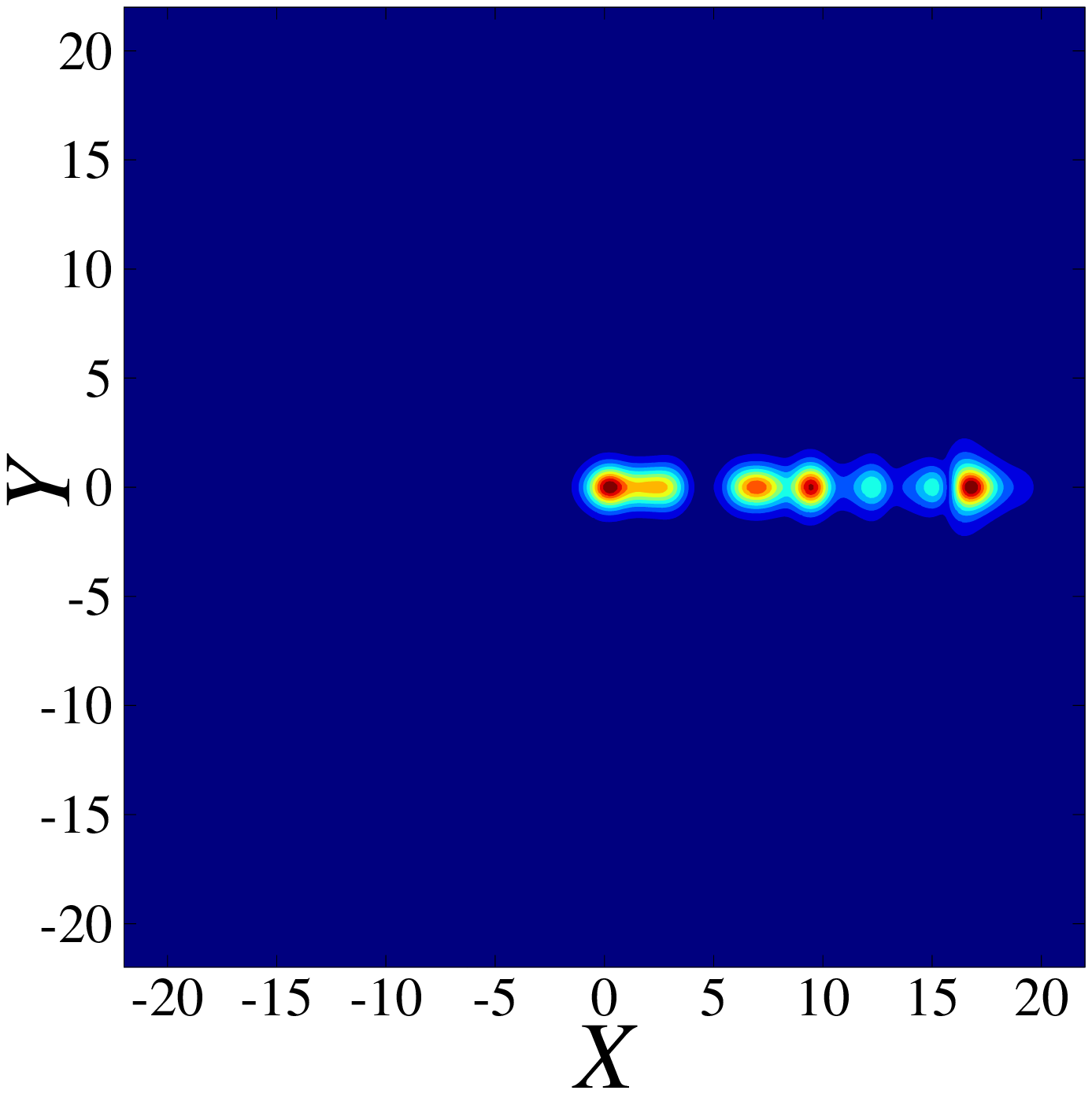}}\vfill
\subfigure[$Z=17.67$]{\label{abssolfondk016theta0piZ30}%
\includegraphics[width=4cm]{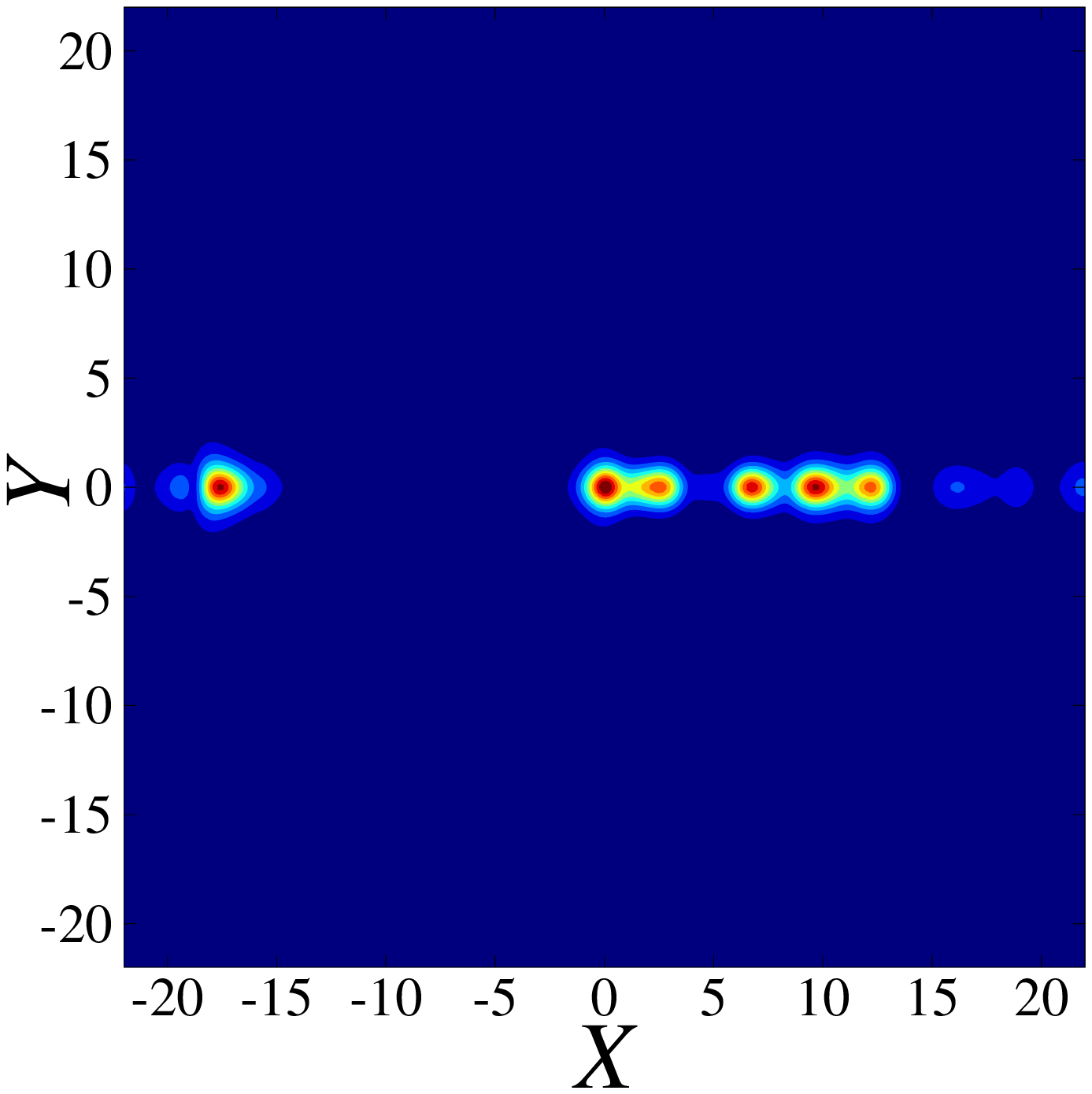}} %
\subfigure[$Z=24.12$]{\label{abssolfondk016theta0piZ45}%
\includegraphics[width=4cm]{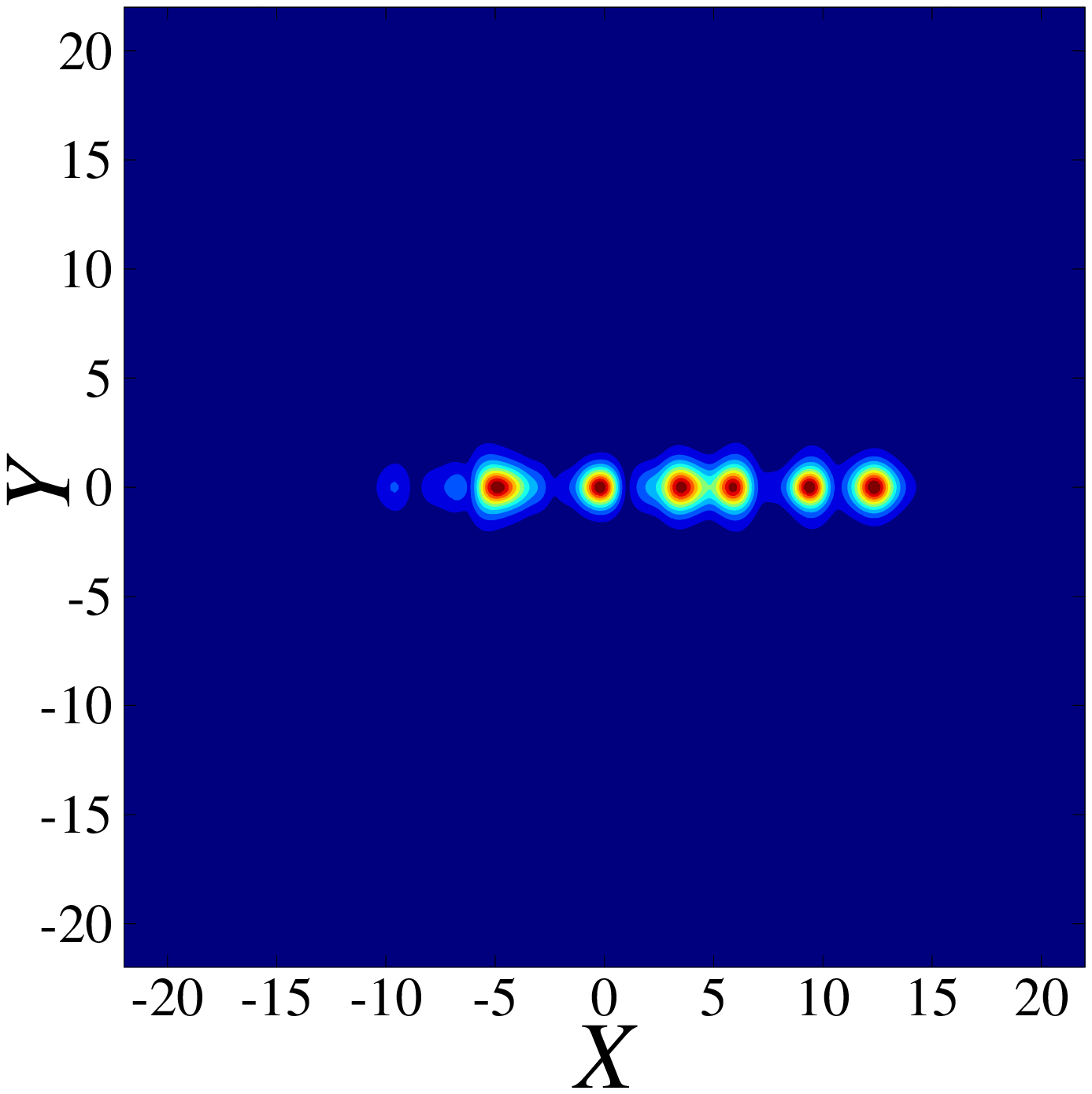}}
\end{center}
\caption{(Color online) The same as in Figs. \protect\ref{fond3} and \protect
\ref{fond4}, but for $k_{0}=1.694$. In this case, the kicked soliton
eventually creates an arrayed set of five solitons (in panel (f), the
additional freely moving (sixth) soliton hits the array from the opposite
direction, completing its round trip in the system).}
\label{fond6}
\end{figure}

On the other hand, the smaller kick can initiate the creation of an arrayed
pattern. This outcome of the evolution is shown in Fig. \ref{fond6}, where
the array of five solitons is created, starting with the soliton initially
kicked by $k_{0}=1.694$, in addition to which a free soliton keeps moving as
a quasi-particle (cf. Ref. \cite{Kominis}), until it collides with the array
from the opposite direction, due to the periodic boundary conditions along $%
X $, and is subsequently absorbed by the array (the collision is displayed
in panel \ref{fond6}(f), where an additional soliton is observed at $X<0$).
This dynamical scenario is additionally illustrated below by Fig. \ref%
{solfon1693}.

\begin{figure}[th]
\begin{center}
\includegraphics[width=5cm]{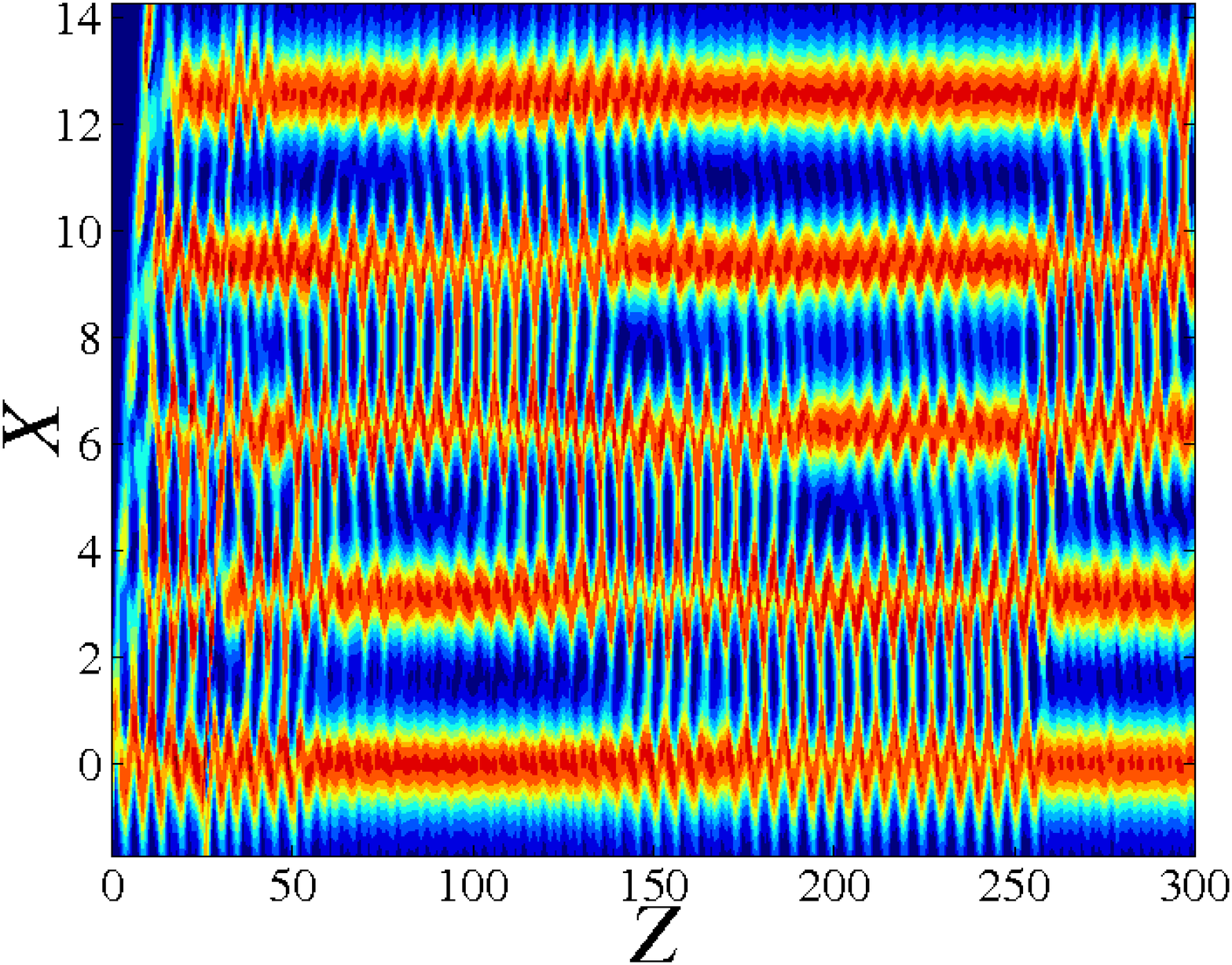}
\end{center}
\caption{(Color online) Evolution of the local amplitude, $|u(X,Z)|$, in the
cross-section plane, $Y=0$, in the same case as in Fig. \protect\ref{fond6} (%
$k_{0}=1.694$ and $\protect\theta =0$).}
\label{fond7}
\end{figure}

The emerging array remains in an excited state, featuring localized density
waves running across it, as shown, on a much longer scale of $Z$, by means
of the cross-section picture in Fig. \ref{fond7}). It is worthy to note that
the wave is reflected from the last pinned soliton.
Such localized density perturbations propagating through a chain of pinned
solitons are similar to the so-called `` super-fluxons", which were
investigated experimentally and theoretically in arrays of fluxons
(topological solitons, representing magnetic-flux quanta) pinned in a long
Josephson junction with a periodic lattice of local defects \cite{super}, as
well as in an array of mutually repelling solitons forming a Newton's cradle
in a two-component model of binary Bose-Einstein condensates \cite{cradle}.

\subsection{The dependence of the outcome of the evolution on the strength
of the initial kick}

Results of the systematic analysis of the model are summarized in Fig. \ref%
{evosol}, where the number of solitons in the stable arrayed patterns
established by the end of the simulation, is plotted versus the initial kick
$k_{0}$ for $\theta =0$,
where intervals of the values of $k_{0}$ corresponding to constant numbers
of the solitons are adduced.


\begin{figure}[th]
\begin{center}
\includegraphics[width=5cm]{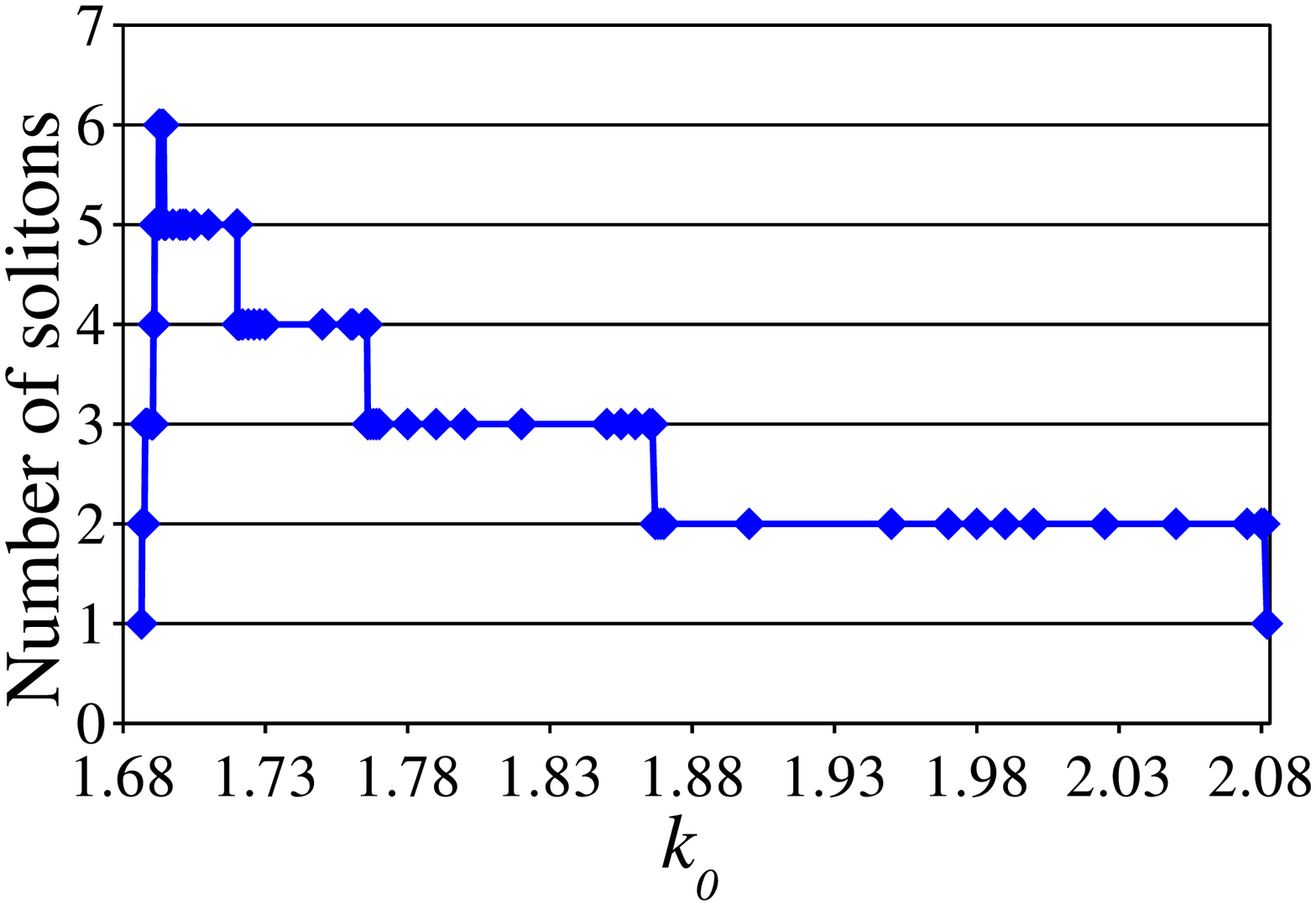}
\end{center}
\caption{(Color online)The number of solitons in the established pattern
versus the kick's strength, $k_{0}$, at $\protect\theta =0$. In narrow gaps
between intervals presented in the graphic, the number cannot be defined
exactly, as it changes there by $1$}
\label{evosol}
\end{figure}

\begin{figure}[th]
\subfigure[$Z=2.60$]{%
\includegraphics[width=4cm]{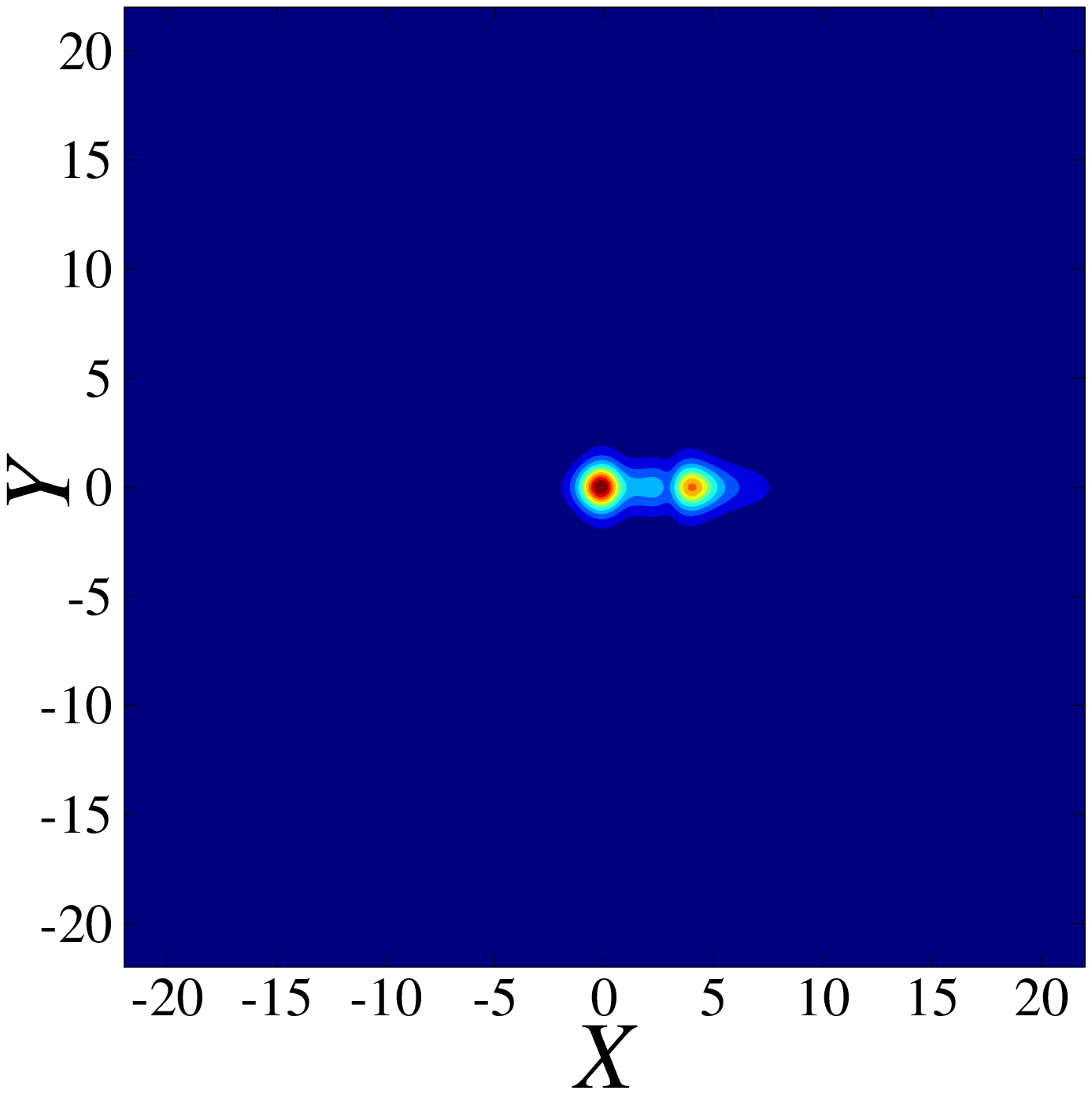}} %
\subfigure[$Z=12.47$]{%
\includegraphics[width=4cm]{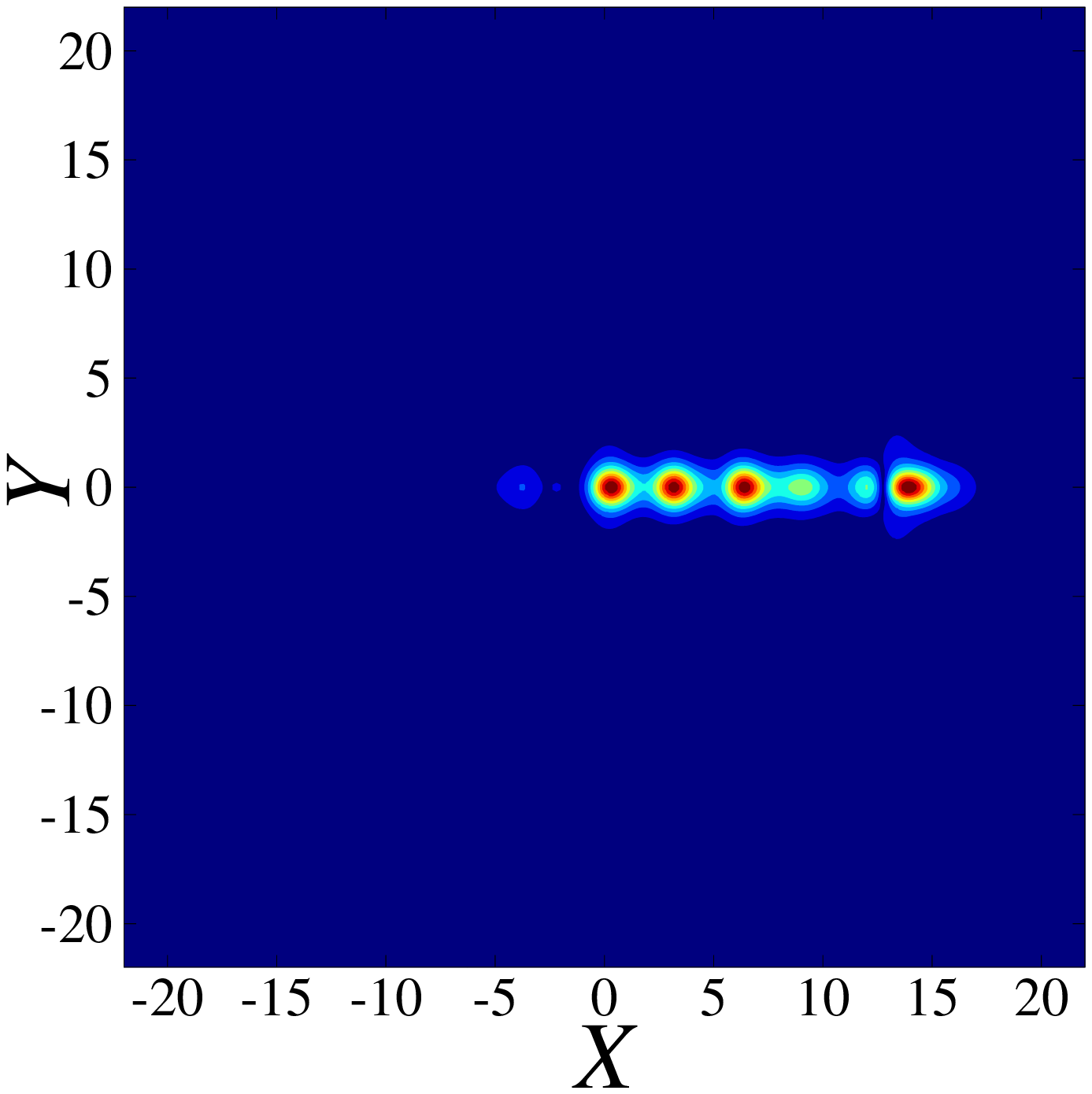}}\vfill
\subfigure[$Z=23.09$]{%
\includegraphics[width=4cm]{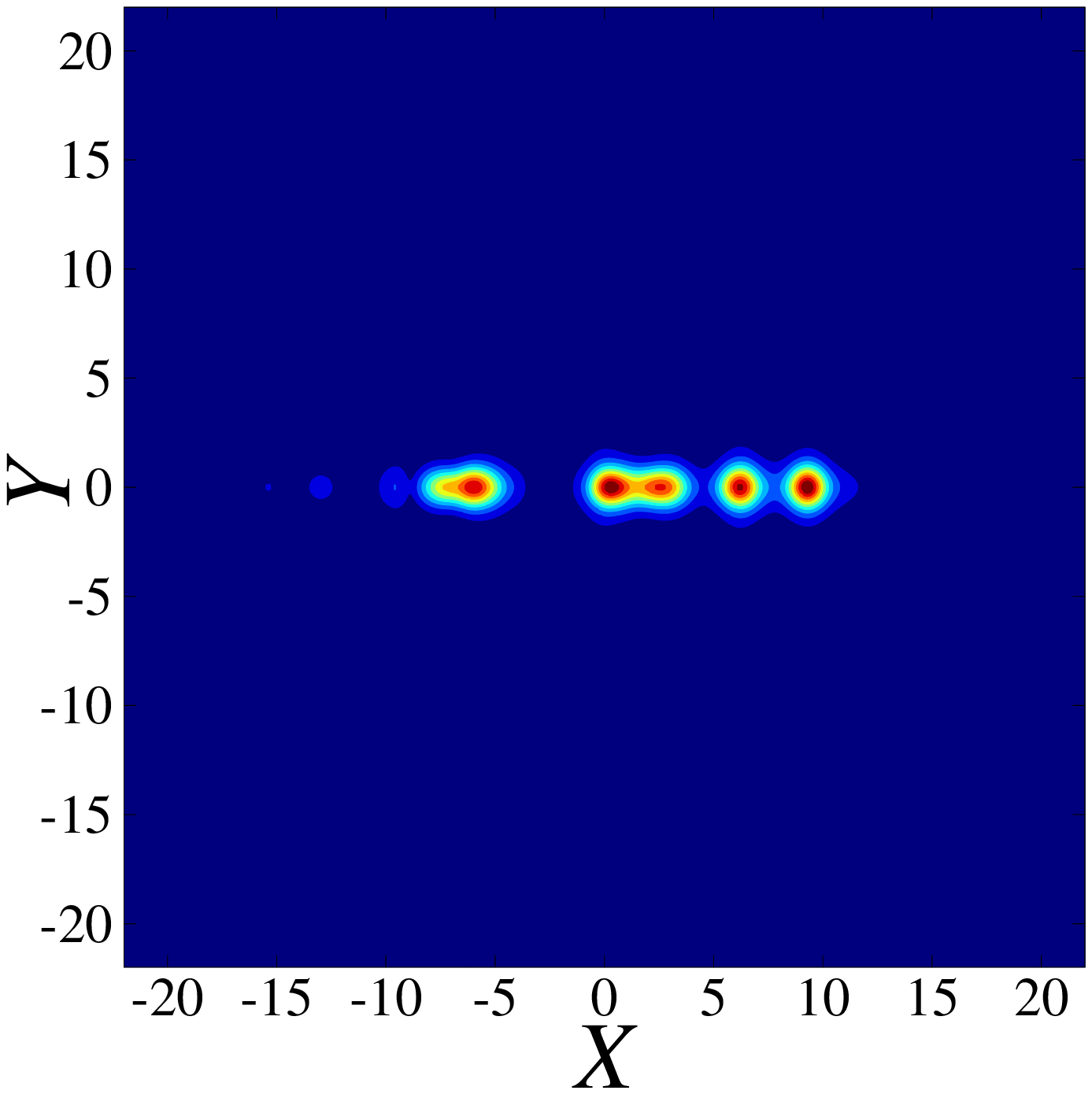}} %
\subfigure[$Z=299.68$]{%
\includegraphics[width=4cm]{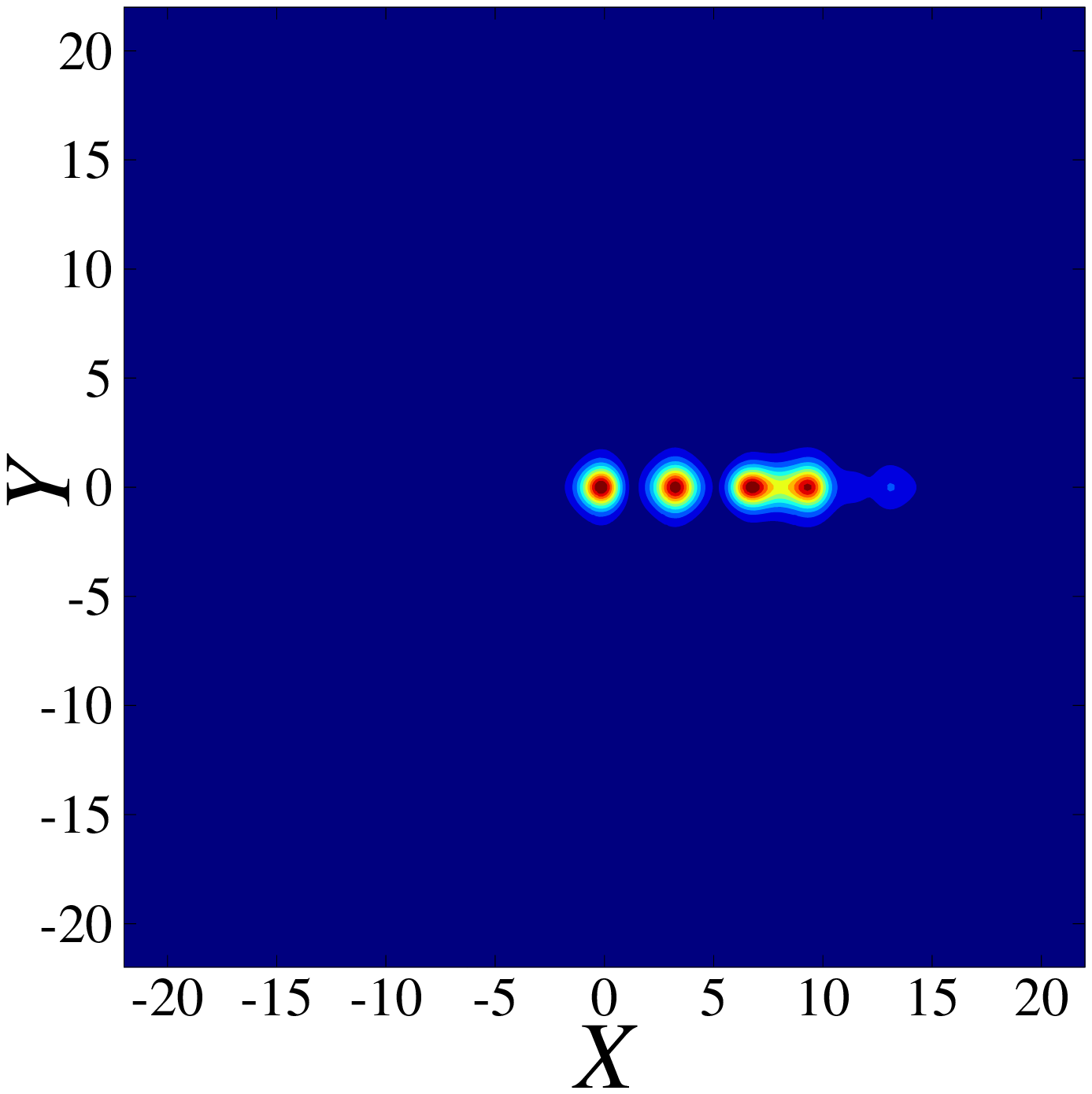}}
\caption{(Color online) The same as in Fig. \protect\ref{fond6}, but for $%
k_{0}=1.705$. In this case, a free soliton splits away from the quiescent
array, then hits it from the other side, and gets absorbed by it.}
\label{fond_cb_k0_17_theta_0}
\end{figure}

\begin{figure}[th]
\begin{center}
\includegraphics[width=5cm]{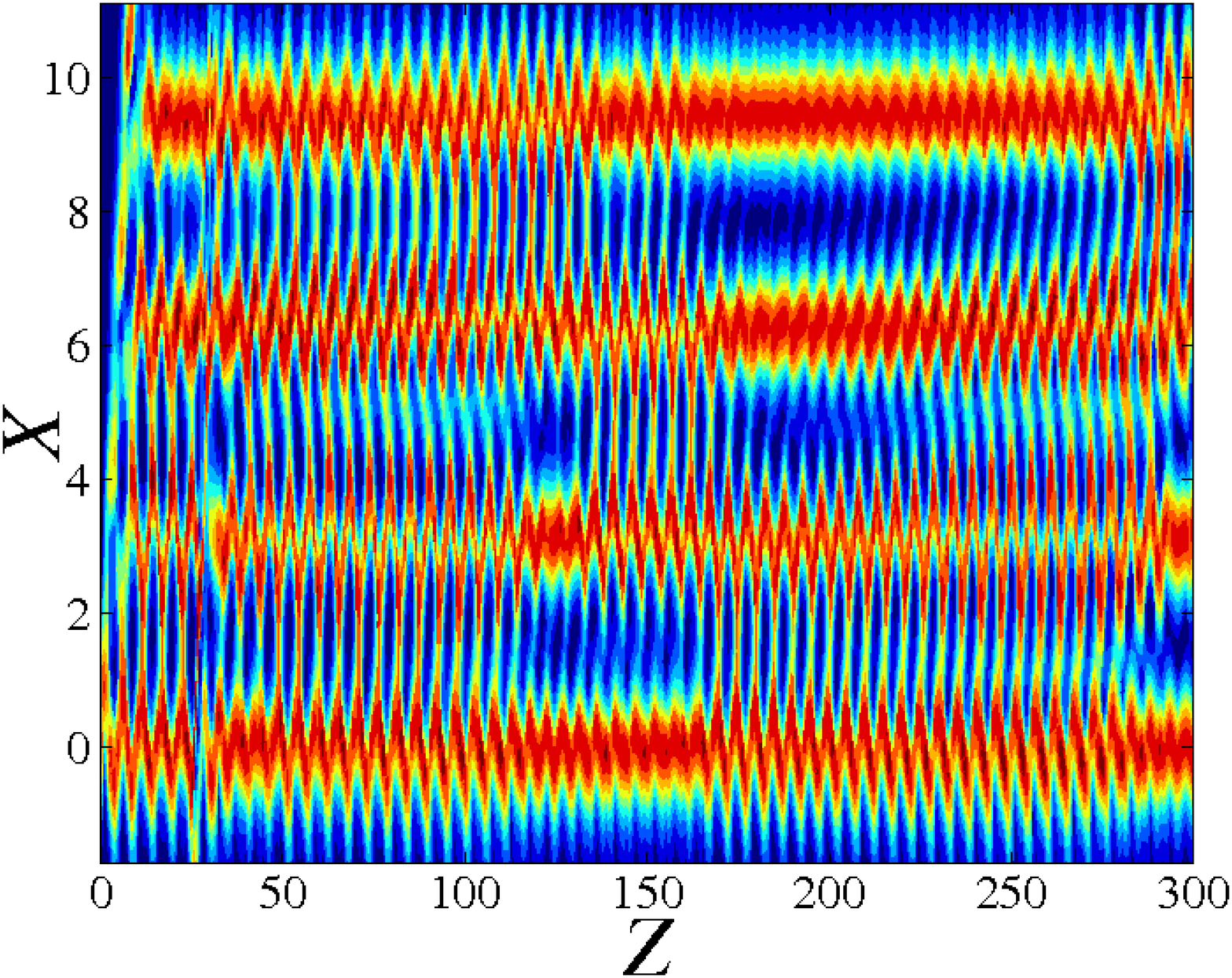}
\end{center}
\caption{(Color online) The top view of the same dynamical picture as in
Fig. \protect\ref{fond_cb_k0_17_theta_0}. }
\label{fond_cb_k0_17_ny}
\end{figure}

In case the free soliton collides with the quiescent array after performing
the round trip in the system with the periodic boundary conditions, see the
example above in Fig. \ref{fond6}, and an additional one (for four solitons)
in Figs. \ref{fond_cb_k0_17_theta_0} and \ref{fond_cb_k0_17_ny}, the number
of solitons was counted just before the first such collision. Otherwise, the
number was recorded after any motion in the system would cease.

The results demonstrate that the number of solitons in the established
patterns rapidly increases from 1 at $k_{0}\leq (k_{0})_{\mathrm{thr}%
}\approx 1.6865$ [see Eq. (\ref{thr})] to a maximum of five solitons, plus a
sixth freely moving one, at $k_{0}=1.6927$.
New solitons add to the established pattern according to the scenario
outlined by means of the bullet items in the previous subsection. At $%
k_{0}>1.6927$, the soliton number decreases by steps with an increasing
length of the corresponding intervals of the kick's strength, see Fig. \ref%
{evosol}.

\begin{figure}[th]
\centering
\subfigure[]{\label{absusolfon1693}%
\includegraphics[height=3.5cm]{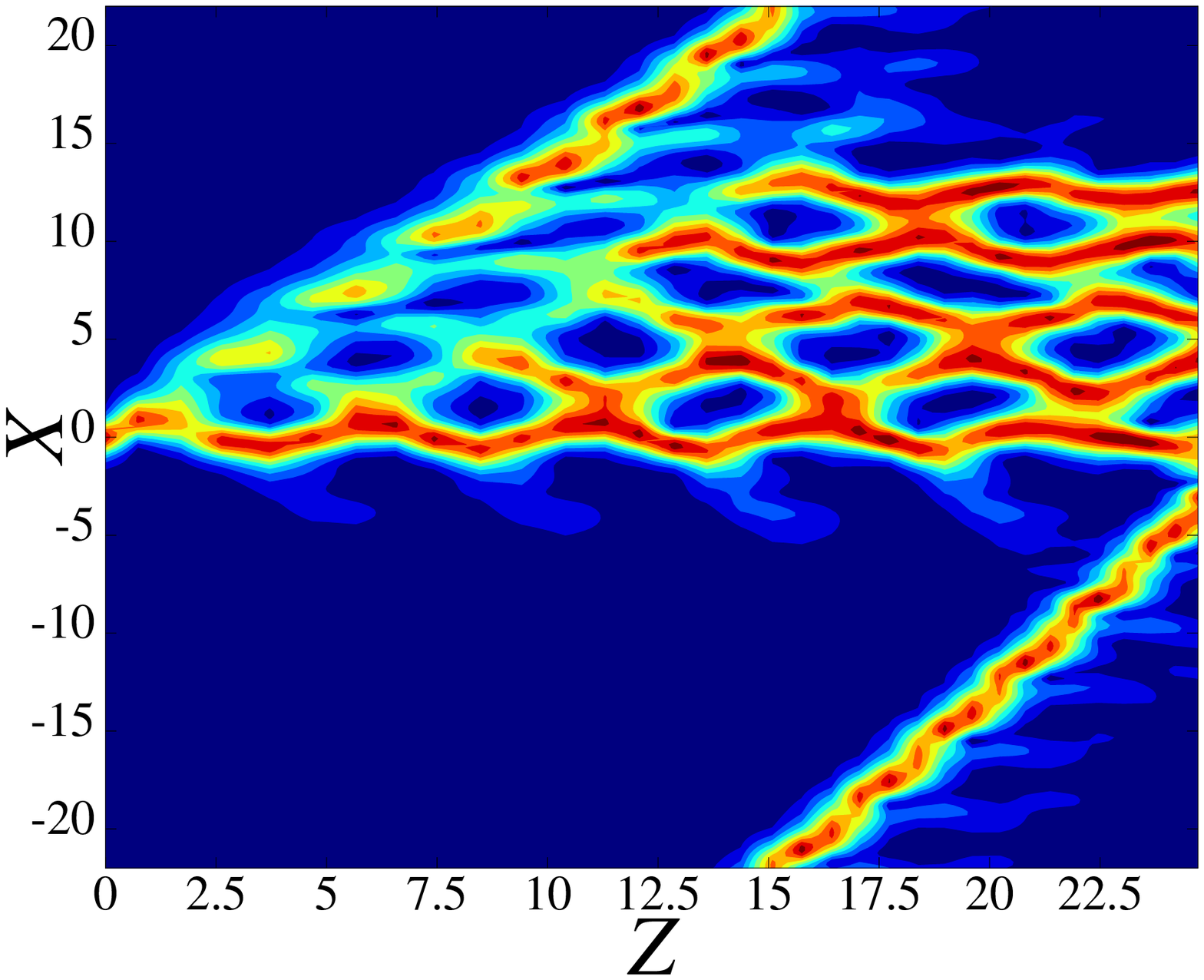}} %
\subfigure[]{\label{energsolfon1693}%
\includegraphics[height=3.5cm]{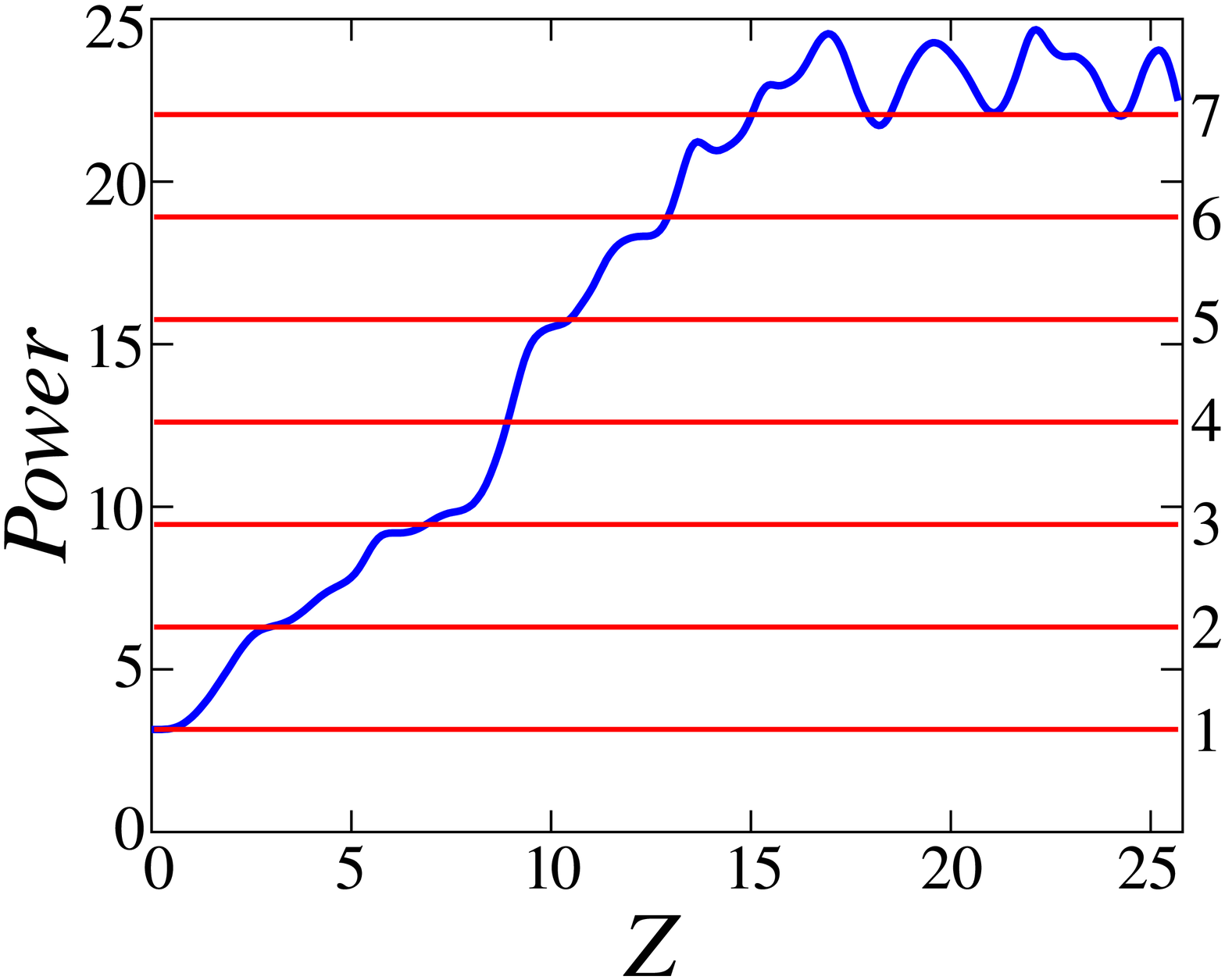}}
\caption{(Color online) (a): The field amplitude, $\left\vert u\left(
X,Z\right) \right\vert $, in the cross section $Y=0$, and (b) the evolution
of the total power, for the kick's strength $k_{0}=1.693$ at $\protect\theta %
=0$, which leads to the establishment of the six-soliton pattern. }
\label{solfon1693}
\end{figure}

As mentioned above, the largest number of six solitons is attained at $%
1.6927<k_{0}<1.6942$. In addition to Figs. \ref{fond6} and \ref{fond7}, this
situation is illustrated by Fig. \ref{absusolfon1693}), where the averaged
total power is $P=23$, see panel \ref{solfon1693}(b). The horizontal
reference lines in this figure show the power levels corresponding to a
single quiescent soliton (recall $P_{\mathrm{sol}}\approx 3.15$), multiplied
by factors from $1$ to $7$, which demonstrates that the total power of the
six-soliton complex exceeds the seven-fold power of the single soliton. This
is due to the fact that the energy of the moving soliton is roughly twice
that of a soliton at rest.

\begin{figure}[th]
\centering
\subfigure[]{\label{abssolfond21}%
\includegraphics[height=3.5cm]{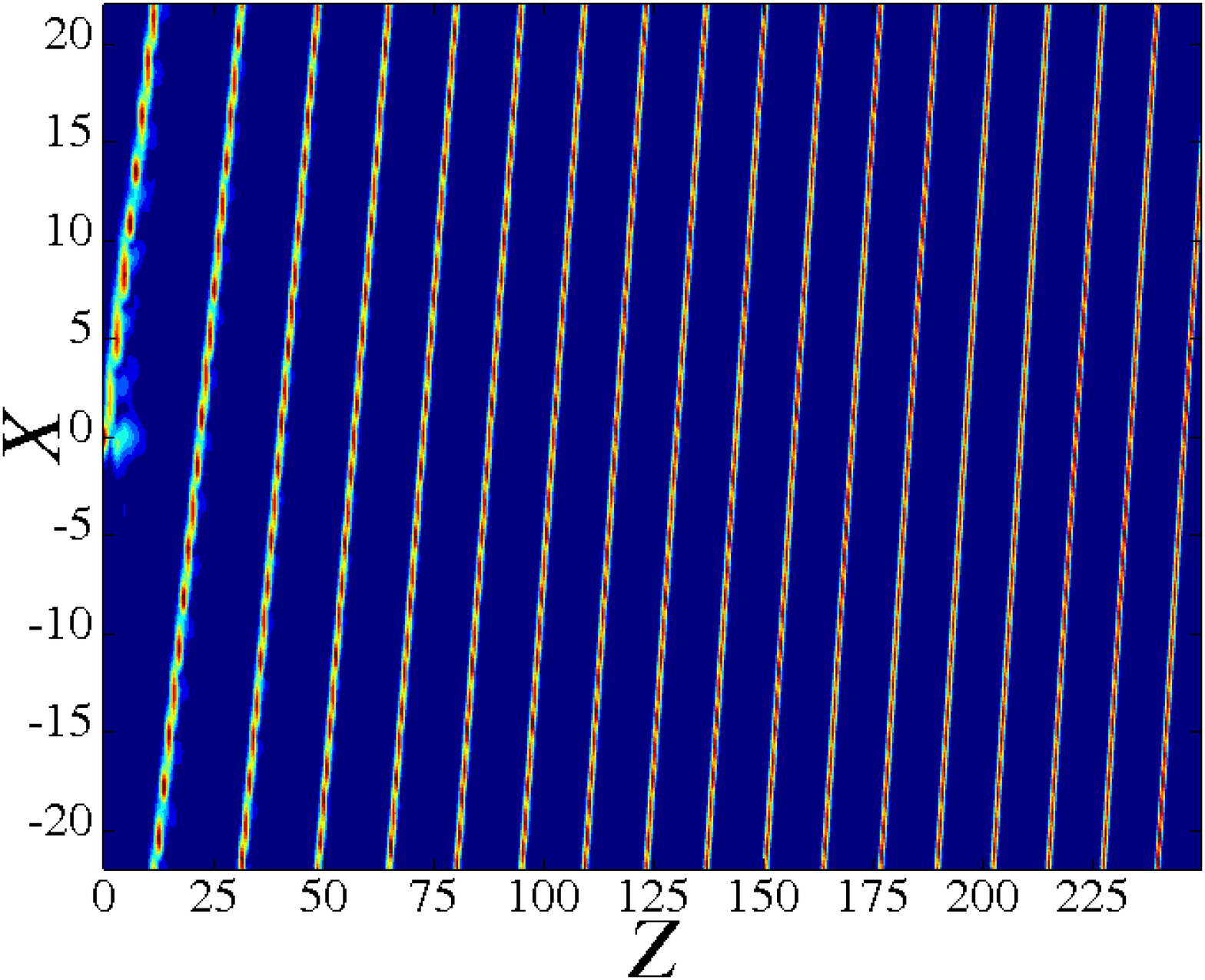}} %
\subfigure[]{\label{vitgsolfond21}%
\includegraphics[height=3.5cm]{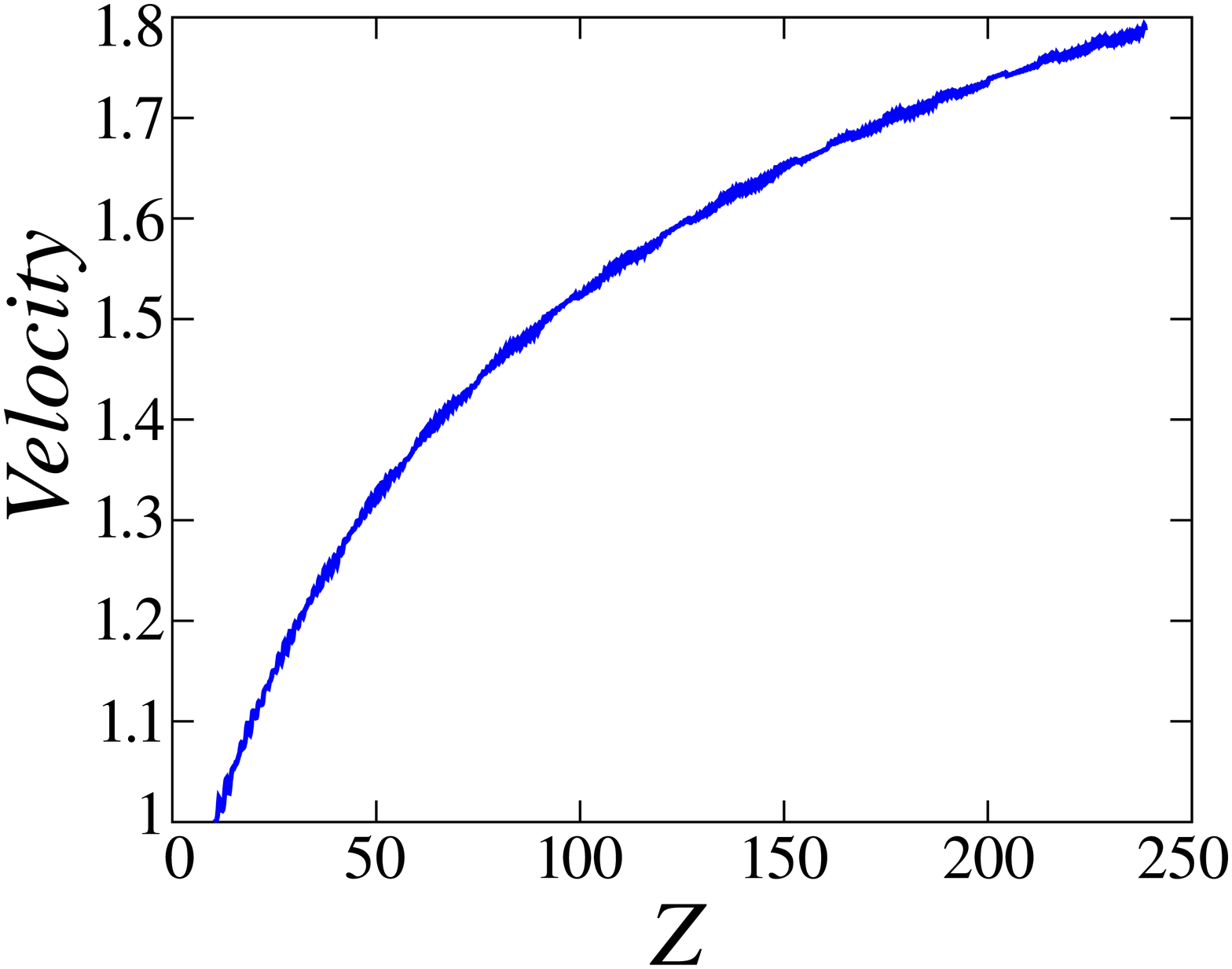}}
\caption{(Color online) (a) The distribution of the field amplitude, $%
\left\vert u\left( X,Z\right) \right\vert $ in cross section $Y=0$, and (b)
the evolution of the velocity of the fundamental soliton kicked by $%
k_{0}=2.1 $ at $\protect\theta =0$.}
\label{solfond21}
\end{figure}

At $k_{0}\geq 2.082$, the kicked soliton moves freely across the simulation
domain, see Fig. \ref{abssolfond21}. In this case, the figure demonstrates
that the soliton's velocity increases, approaching a certain limit value.
The computation of the velocity was performed by means of the Lagrange's
interpolation of the numerical data to accurately identify the soliton's
center. To display the results, small-scale oscillations of the velocity of
the soliton passing the periodic potential (obviously, the velocity is
largest and smallest when the solitons traverses the bottom and top points
of the potential, respectively) have been smoothed down by averaging the
dependence over about fifteen periods.

\subsection{The evolution initiated by an oblique kick}

The application of the kick under an angle to the lattice, i.e., with $%
\theta \neq 0$ [see Eq. (\ref{k})], was considered too. The results are
summarized in Table \ref{tableevolsolfondpi4} for $\theta =\pi /4$ (the kick
oriented along the diagonal), and, below, in Table \ref{tableevolsolfondpi8}
for $\theta =\pi /8$.

\begin{table}[th]
\begin{center}
\begin{tabular}{|c||c|}
\hline
Number of solitons & Range of $k_0$ \\ \hline
$1$ & $k_0\in[0,2.355]$ \\ \hline
$0$ & $k_0\in[2.36,4.337]$ \\ \hline
$1$ & $k_0\in[4.563,10]$ \\ \hline
\end{tabular}%
\end{center}
\caption{The number of solitons in the established pattern versus the kick's
strength, $k_{0}$, at $\protect\theta =\protect\pi /4$.}
\label{tableevolsolfondpi4}
\end{table}

For $\theta =\pi /4$, the initial soliton remains pinned at
\begin{equation}
k_{0}\leq \left( k_{0}\right) _{\mathrm{thr}}\left( \theta =\pi /4\right)
=2.166,  \label{pi/4}
\end{equation}
and it is destroyed at $2.25<k_{0}<4.337$. At $k_{0}>4.337$, the single
soliton survives, moving freely along the diagonal direction. Thus, the
final number of the solitons in this case is $1$ or $0$ (see Table \ref%
{tableevolsolfondpi4}), and no new solitons are generated. As concerns the
comparison with the analytical prediction (\ref{analyt}), it yields $\left(
k_{0}\right) _{\mathrm{thr}}\left( \theta =\pi /4\right) \approx 1.87$,
which, as well as in the case of $\theta =0$, is somewhat lower than its
numerical counterpart.

\begin{figure}[th]
\centering
\subfigure[]{\label{abssolfond100theta025pi}%
\includegraphics[width=4cm]{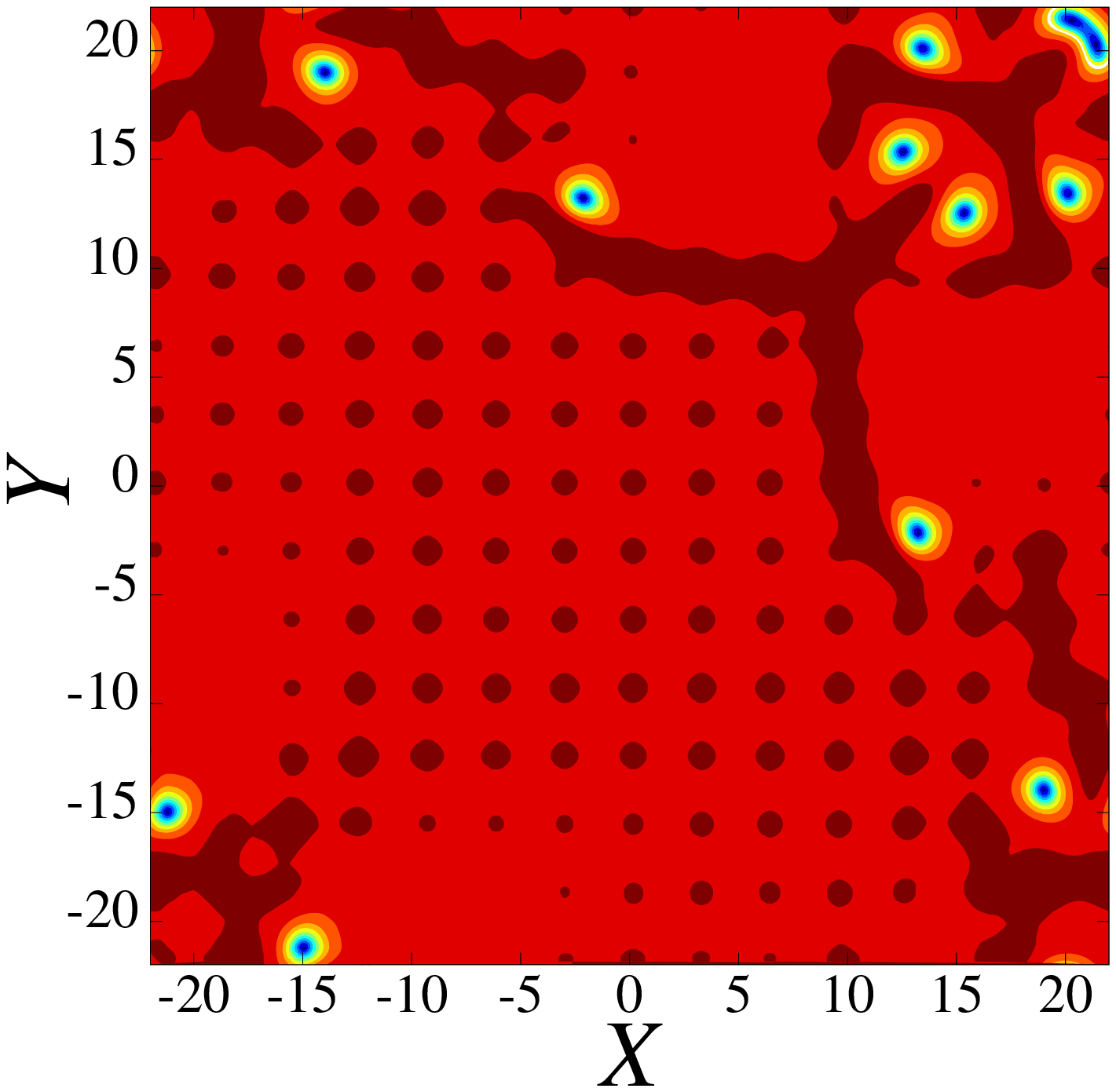}}%
\subfigure[]{\label{phasesolfond100theta025pi}%
\includegraphics[width=4cm]{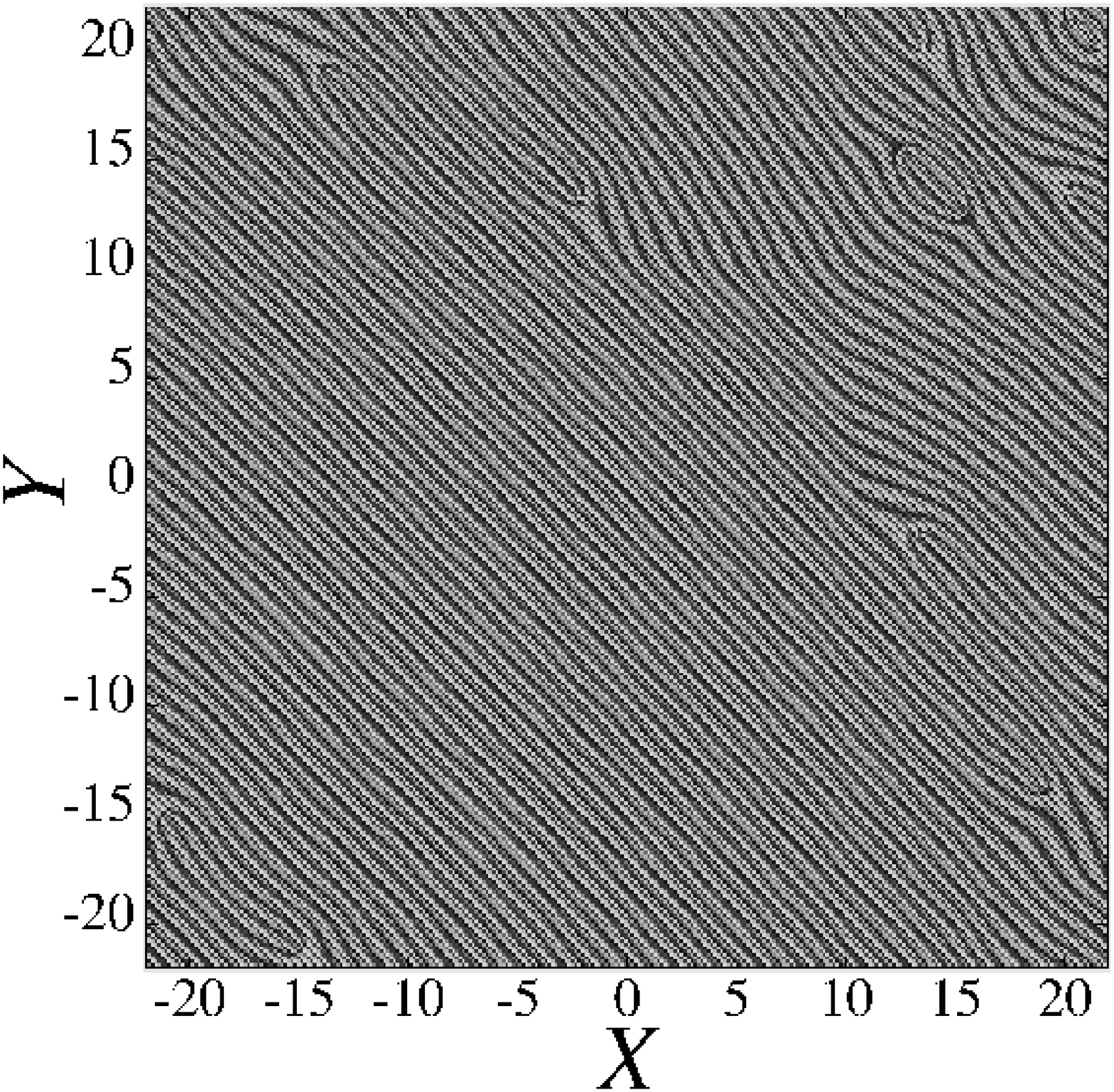}}%
\vfill
\caption{(Color online) The distribution of the field amplitude, $\left\vert
u\left( X,Y\right) \right\vert $ (a) and phase (b) in the pattern produced
by kicking the fundamental soliton in the diagonal direction ($\protect%
\theta =\protect\pi /4$) very hard, with $k_{0}=100$. The picture
corresponds to evolution distance $Z=49.98$.}
\label{solfond100theta025pi}
\end{figure}

The kick with much larger values of $k_{0}$ (one or two orders of magnitude
higher than in Table \ref{tableevolsolfondpi4}) causes the generation of
dark-soliton structures, supported by a nonzero background filling the
entire domain (the total power may then exceed that of the single soliton by
a factor $\sim 1000$). Inspection of Fig. \ref{abssolfond100theta025pi}
reveals stable holes in the continuous background, whose centers coincide
with phase singularities (see Fig. \ref{phasesolfond100theta025pi}). These
holes represent vortices supported by the finite background.

The creation of new solitons is possible in the case of the oblique kick
with $\theta =\pi /8$. In this case, the new solitons may be oriented along
either axis, $X$ or $Y$, as indicated in Table \ref{tableevolsolfondpi8}
(the total count includes the obliquely moving originally kicked soliton).

\begin{table}[th]
\begin{center}
\begin{tabular}{|c|c|c||c|}
\hline
\multicolumn{3}{|c||}{Number of solitons} & %
\raisebox{-1.5ex}[0pt][0pt]{Range of $k_0$} \\ \cline{1-3}
{\small Total} & {\small Along $X$} & {\small Along $Y$} &  \\ \hline
$1$ & $0$ & $0$ & $k_0\in[0,1.816]$ \\ \hline
$3$ & $2$ & $0$ & $k_0=1.974$ \\ \hline
$2$ & $1$ & $0$ & $k_0=2.1$ \\ \hline
$1$ & $0$ & $0$ & $k_0\in[2.224,4.569]$ \\ \hline
$0$ & $0$ & $0$ & $k_0\in[4.816,5.804]$ \\ \hline
$1$ & $0$ & $0$ & $k_0\in[6.05,\infty)$ \\ \hline
\end{tabular}%
\\[0pt]
\end{center}
\caption{The same as in Table \protect\ref{tableevolsolfondpi4}, but for the
oblique kick oriented under angle $\protect\theta =\protect\pi /8$.}
\label{tableevolsolfondpi8}
\end{table}

Further, for the kick applied at angle $\theta =\pi /8$, the simulations
demonstrate that the\ kicked soliton remains pinned at
\begin{equation}
k_{0}\leq \left( k_{0}\right) _{\mathrm{thr}}\left( \theta =\pi /8\right)
=1.816,  \label{pi/8}
\end{equation}
while analytical approximation (\ref{analyt}) for the same case predicts $%
\left( k_{0}\right) _{\mathrm{thr}}\left( \theta =\pi /8\right) \approx 1.54$%
. The creation of new solitons occurs above the threshold, similar to the
case of $\theta =0$ and in contrast with $\theta =\pi /4$. The largest
three-soliton pattern is created at $k_{0}=1.974$. It is composed of two
oscillating pinned solitons and a freely moving one, as shown in Fig. \ref%
{absusolfond1974}. At $k_{0}=2.1$, the dynamics again amounts to the motion
of a single soliton. Further, the simulations demonstrate that, in all the
cases of the free motion of the single soliton, it runs strictly along the $%
X $-axis, despite the fact that the initial kick was oblique.

\begin{figure}[th]
\centering
\includegraphics[width=5cm]{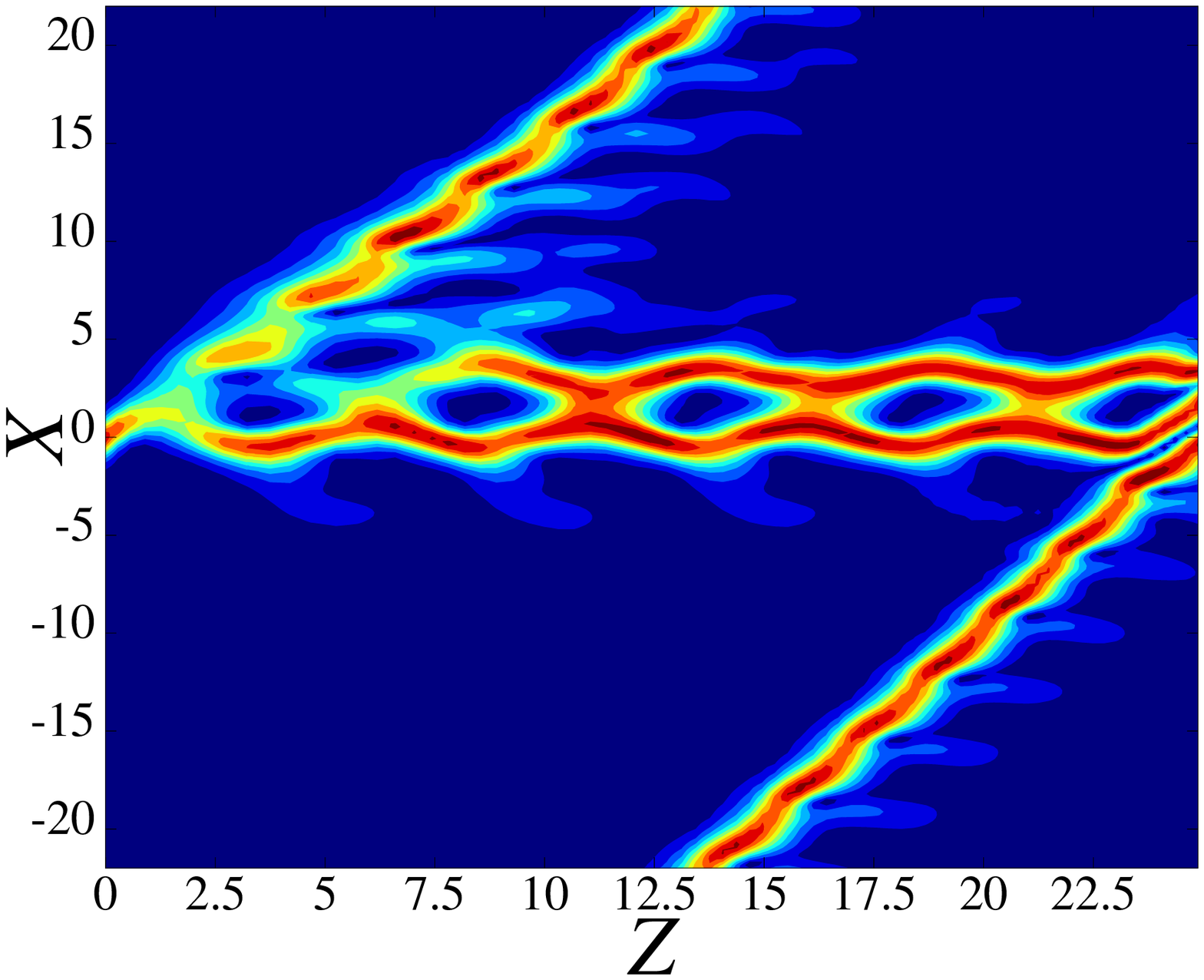}
\caption{(Color online) The local field amplitude, $\left\vert u\left(
X,Z\right) \right\vert $, in the cross section $Y=0$, for $k_{0}=1.9743$ and
$\protect\theta =\protect\pi /8$.}
\label{absusolfond1974}
\end{figure}

\begin{figure}[th]
\centering
\includegraphics[width=4cm]{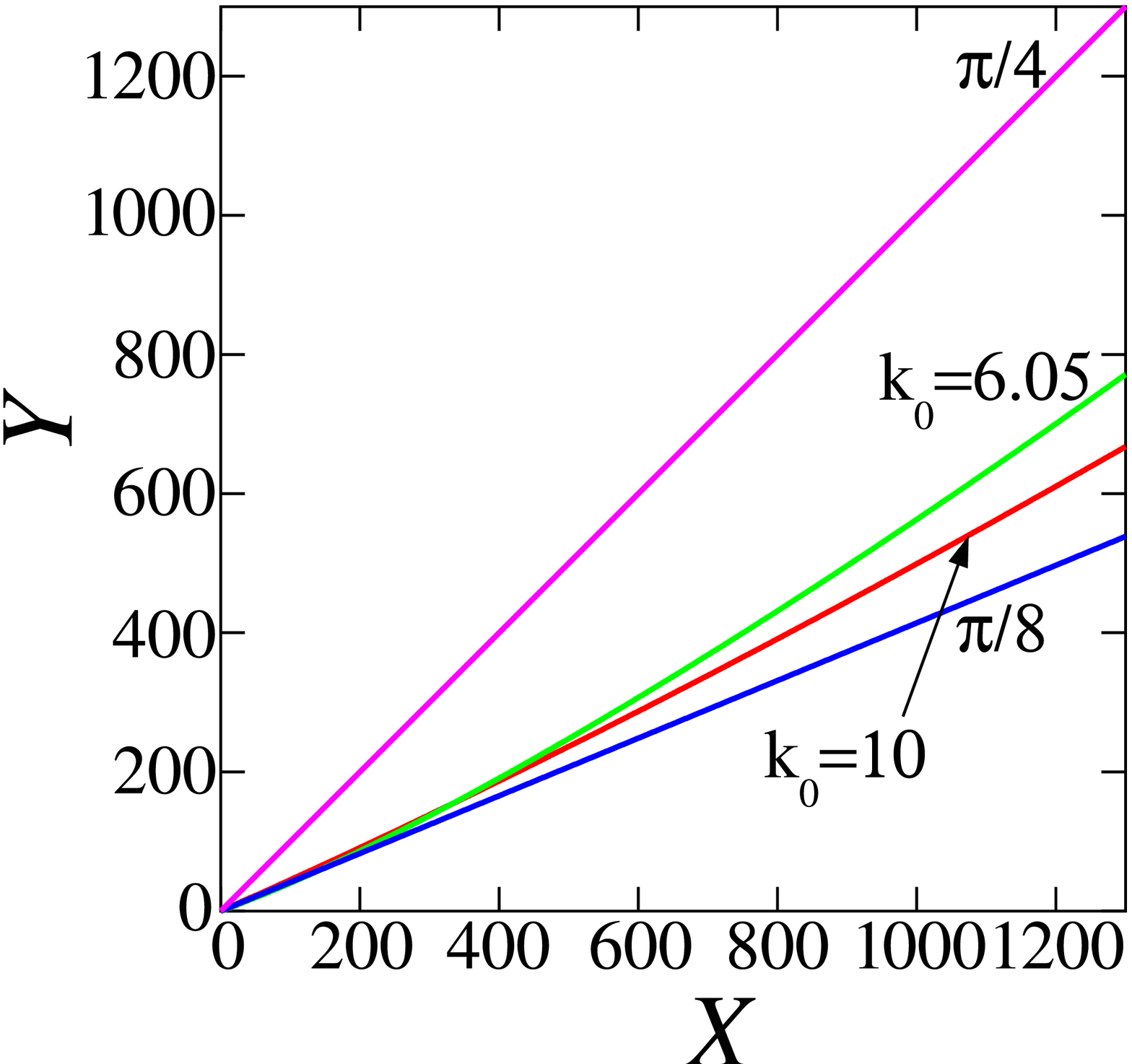}
\caption{(Color online) Examples of curvilinear trajectories of the soliton
for $\protect\theta =\protect\pi /8$ and two large values of the kick, $%
k_{0}=6.05$ and $10$. For the references, straight lines with sloped $%
\protect\theta ={\protect\pi }/{8}$ and $\protect\theta ={\protect\pi }/{4}$
are displayed too.}
\label{angledplthetapi8}
\end{figure}

A harder kick,
\begin{equation}
4.816<k_{0}<5.804,  \label{destr}
\end{equation}%
destroys the soliton (its power at first increases, as it moves across the
first PN barrier, and then decays to zero). At still higher $k_{0}$, the
soliton survives the kick, but in this case its trajectory is curvilinear in
the plane of $\left( X,Y\right) $as shown in Fig. \ref{angledplthetapi8}.

It is relevant to compare the number of solitons in the patterns generated
by the simulations for $\theta =0$ and $\theta ={\pi }/{8}$. For both cases,
these numbers are presented, as functions of projection $k_{0x}\equiv
k_{0}\cos \theta $ of the kick vector onto the $X$-axis, in Table \ref%
{tableprojete}. It is seen that the dependences of the soliton number on $%
k_{0x}$ are quite similar for $\theta =0$ and $\theta =\pi /8$, barring the
case of the destruction of the soliton ($0$ in the table), which occurs at $%
\theta =\pi /8$, but does not happen for $\theta =0$.

\begin{table}[th]
\begin{center}
\begin{tabular}{|c||c|c|c|}
\hline
Range of $k_{0}$ & Range of the & Number of solitons & Number of solitons \\
& projection $k_{0x}$ & $\theta =0$ & $\theta ={\pi }/{8}$ \\ \hline
$\lbrack 0,1.816]$ & $[0,1.6778]$ & 1 & 1 \\ \hline
$1.974$ & $1.8237$ & 3 & 3 \\ \hline
$2.1$ & $1.9401$ & 2 & 2 \\ \hline
$\lbrack 2.224,4.569]$ & $[2.0547,4.2212]$ & 1 & 1 \\ \hline
$\lbrack 4.816,5.804]$ & $[4.4494,5.3622]$ & 1 & 0 \\ \hline
\end{tabular}%
\\[0pt]
\end{center}
\caption{The number of solitons versus the $k_{0x}$ component of the kick
vector, for orientations $\protect\theta =0$ and $\protect\pi /8$.}
\label{tableprojete}
\end{table}

\begin{figure}[th]
\centering
\includegraphics[width=4cm]{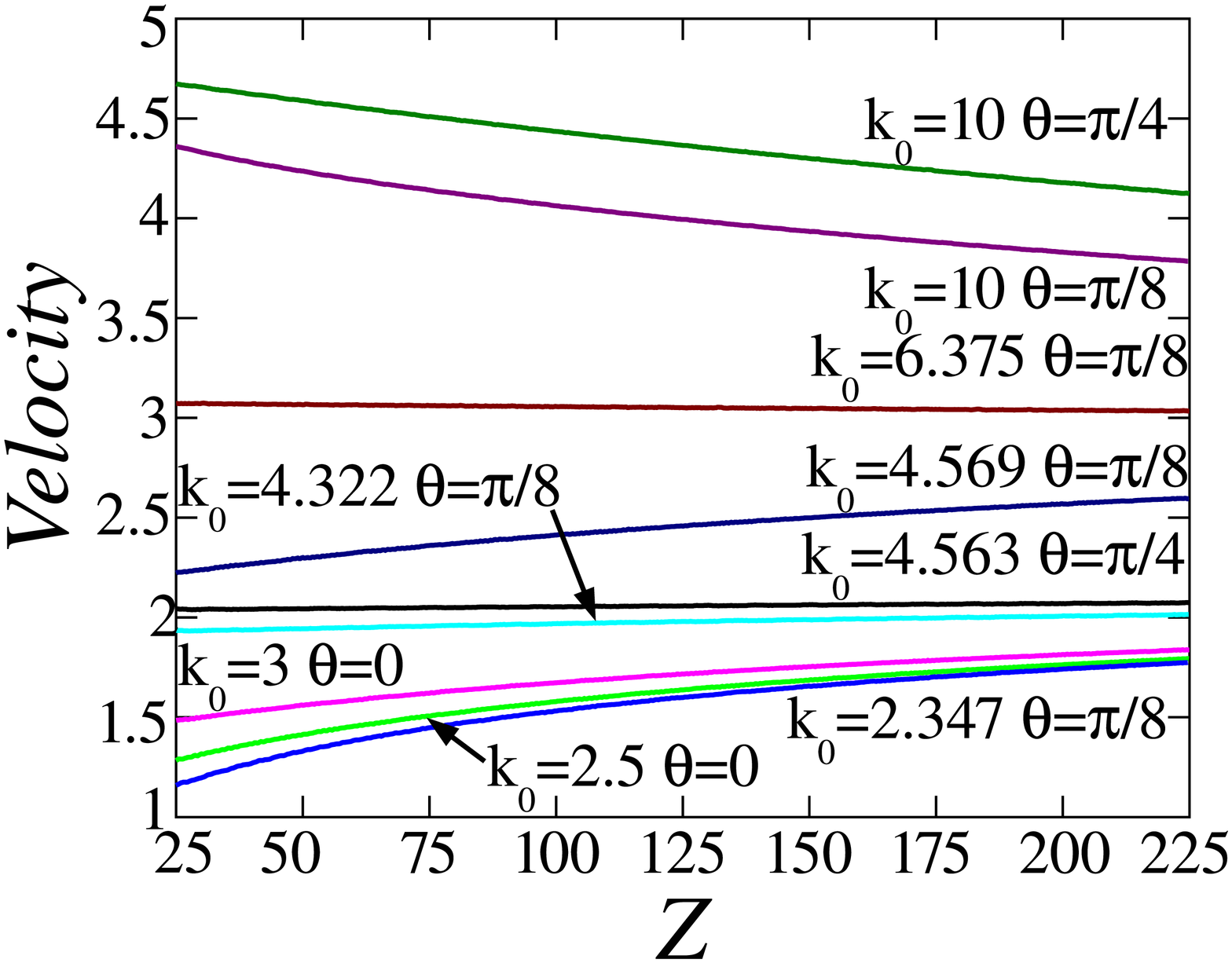}
\caption{(Color online) The soliton's velocity as a function of $Z$ at
various values of $k_{0}$ and $\protect\theta $.}
\label{vitsolfond}
\end{figure}

Further, the evolution of the soliton's velocity for different strengths of
the kick is shown in Fig. \ref{vitsolfond}. Recall that, for $\theta ={\pi }/%
{8}$, there are two domains of values of $k_{0}$ in which the kicked soliton
moves, separated by interval (\ref{destr}) where the soliton is destroyed by
the kick. It is observed that the solitons accelerate and decelerate below
and above the nonexistence interval (\ref{destr}), respectively, but
eventually the velocity approaches a constant value. Moreover, the picture
suggests that, as a result of the long evolution, the velocity is pulled to
either of the two discrete values, $\approx 2$ or $\approx 3$. As said
above, these velocities (in other words, the values of $k_{0}$ which can
directly produce such velocities, see straight horizontal lines in Fig. \ref%
{vitsolfond}) correspond to the single soliton moving along the $X$ axis,
rather than under an angle to it. The conclusion that the system relaxes to
discrete values of the velocity is natural, as the dissipative system should
give rise to a single or several isolated \textit{attractors}, rather than a
continuous family of states with an arbitrary velocity.

A similar behavior is observed for other orientations of the kick, $\theta
=\pi /4$ and $\theta =0$, see the curves labeled by these values of $\theta $
in Fig. \ref{vitsolfond}, and also Fig. \ref{vitgsolfond21}. Again, the
velocity asymptotically approaches the same discrete values, close to $2$
and $3$, with the acceleration or deceleration below and above these values,
respectively. In the case of $\theta =\pi /4$, in the established regime the
free soliton moves in the diagonal direction.

\section{Collisions between moving solitons and standing patterns}

\begin{figure}[th]
\centering
\subfigure[]{\label{abscraddle}%
\includegraphics[width=5cm]{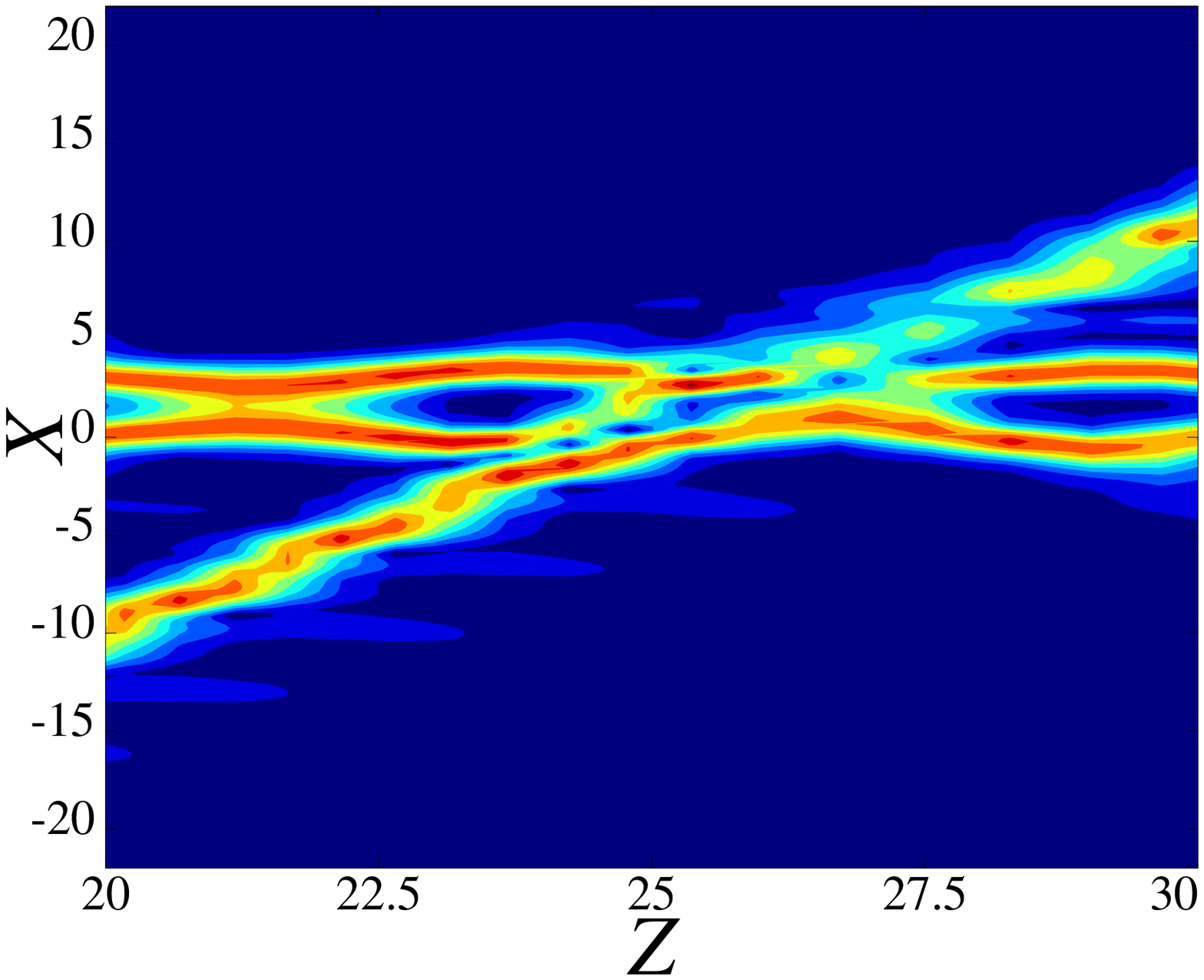}} %
\subfigure[]{\label{absusolfond1765}%
\includegraphics[width=5cm]{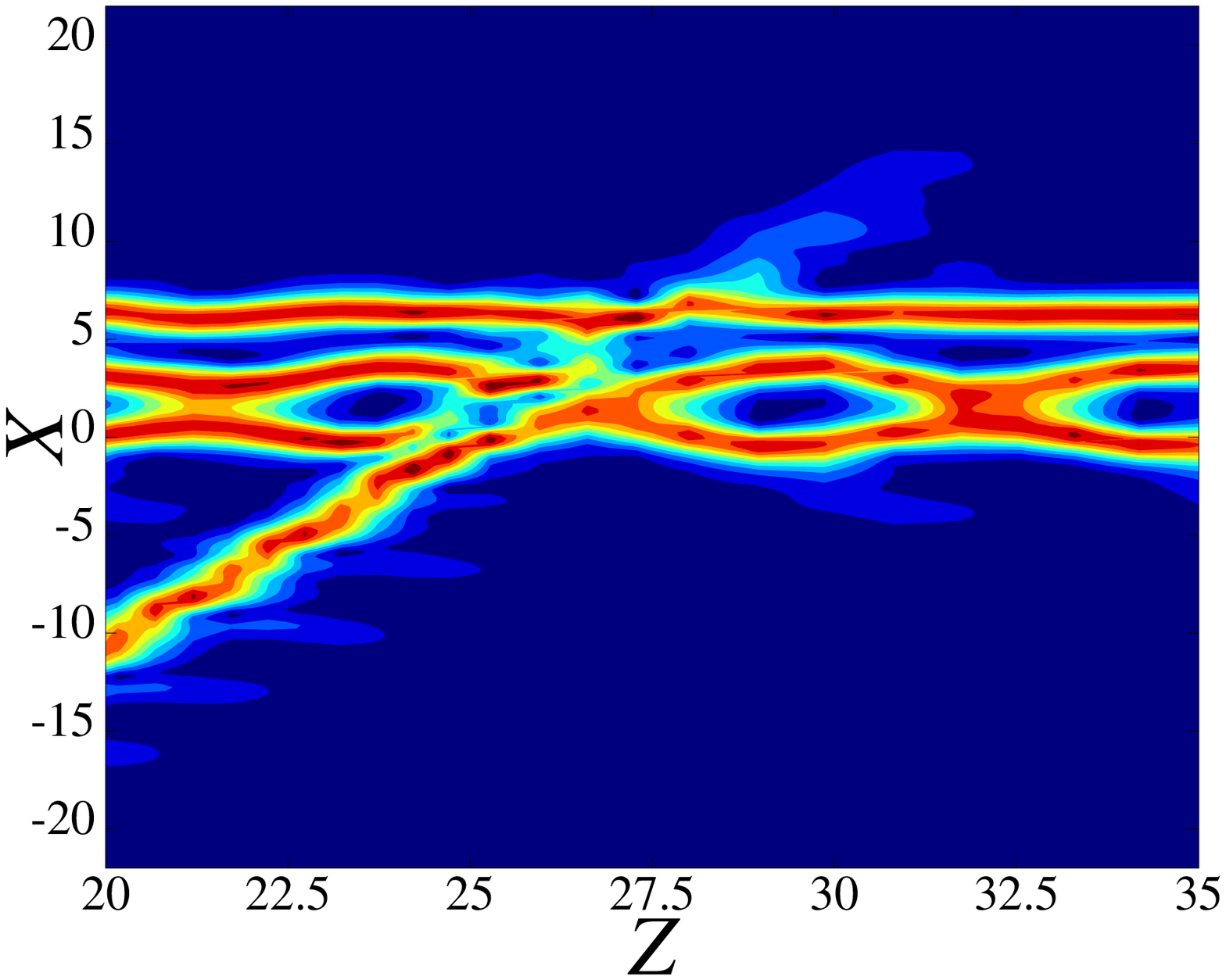}}
\caption{(Color online) Basic types of interactions between the moving
soliton and the quiescent multi-soliton structure: (a) An elastic collision,
at $k_{0}=1.8$ and $\protect\theta =0$; (b) the absorption of the incident
soliton, at $k_{0}=1.765$ and $\protect\theta =0$.}
\label{typecollision}
\end{figure}

One of generic dynamical patters identified above features a standing
multi-soliton structure and a freely moving soliton, which, due to the
periodic boundary conditions, hits the standing structure from the opposite
side, see Figs. \ref{fond6}, \ref{fond_cb_k0_17_theta_0}, \ref{solfon1693},
and \ref{absusolfond1974}. Two distinct scenarios of the ensuing interaction
have been identified in this case, namely, the elastic collision, with the
incident soliton effectively passing the quiescent structure (via a
mechanism resembling the Newton's cradle, cf. Ref. \cite{cradle}) and
reappearing with the original direction and velocity of the motion, and
absorption of the free soliton by the structure, see Figs. \ref{abscraddle}
and \ref{absusolfond1765}, respectively.

\begin{figure}[th]
\centering
\includegraphics[width=5cm]{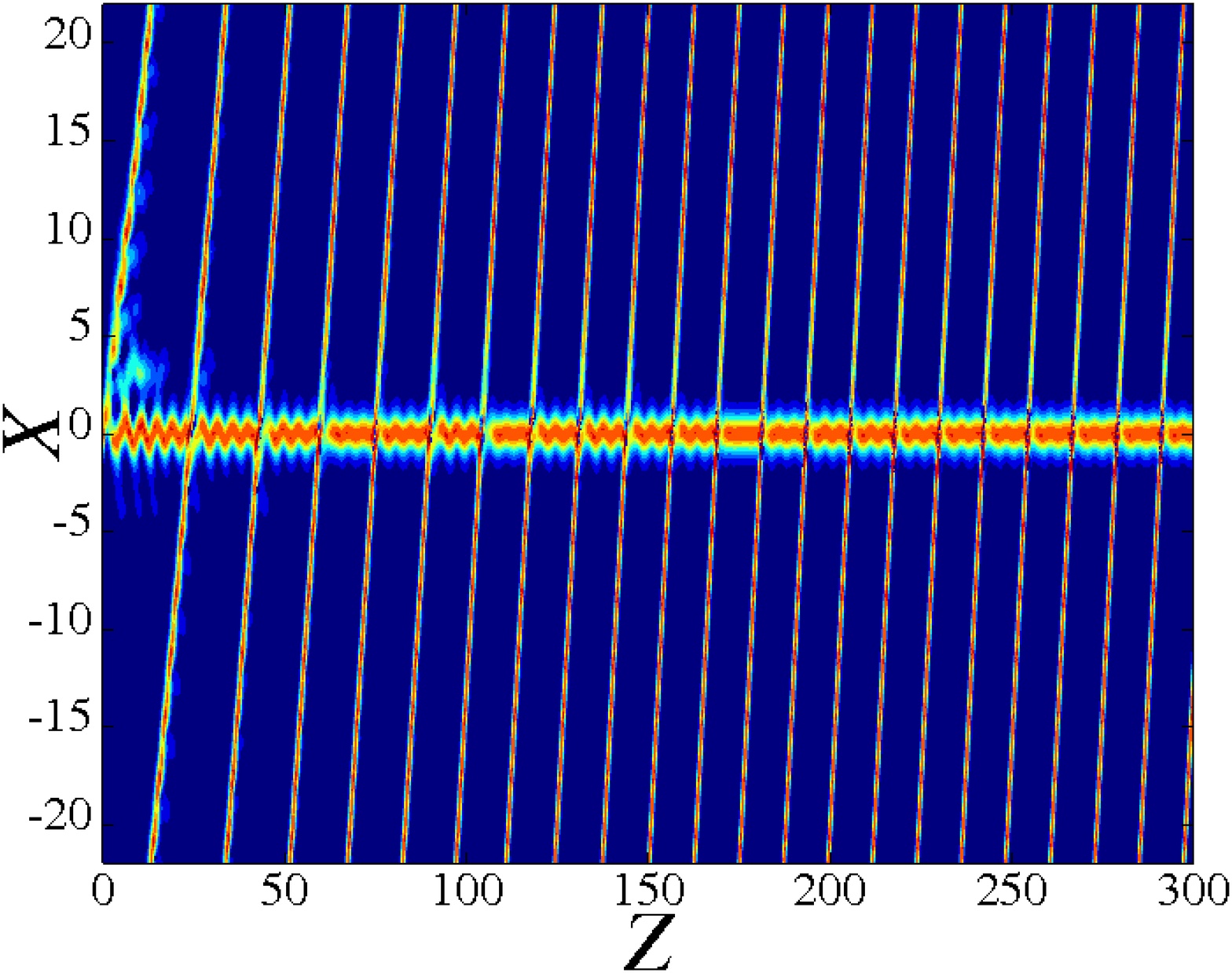}
\caption{(Color online) An example of periodic elastic collisions according
to the Newton's-cradle scenario, at $k_{0}=1.867$ and $\protect\theta =0$.}
\label{absusolfond1867}
\end{figure}

More complex interaction scenarios were observed too, with several elastic
or quasi-elastic collisions that end up with the eventual absorption of the
free soliton. Outcomes of the collisions are summarized in Table \ref%
{domainecollisionk0solfond} (the range of $k_{0}>2.081$ is not shown in the
table, as only the single soliton exists in that case). At $1.766<k_{0}<1.86$%
, several elastic collisions, from $1$ to $5$, the number of which
alternates in an apparently random fashion, are observed before the
absorption is registered. This scenario is labeled `` Newton's cradle with
damping" in Table \ref{domainecollisionk0solfond}. At $k_{0}>1.86$, the
collision is elastic and persists to occur periodically, as in the case of
the ordinary Newton's cradle (so named too in Table \ref%
{domainecollisionk0solfond}), see Fig. \ref{absusolfond1867}.

\begin{table}[th]
\begin{center}
\begin{tabular}{|c||c|}
\hline
Collision type & Range of $k_0$ \\ \hline
absorption & $k_0=1.6879$ \\ \hline
Newton's cradle with damping & $k_0=1.6909$ \\ \hline
absorption & $k_0=1.692$ \\ \hline
complex & $k_0=1.693$ \\ \hline
absorption & $k_0\in[1.695,1.765]$ \\ \hline
Newton's cradle with damping & $k_0\in[1.766,1.86]$ \\ \hline
Newton's cradle & $k_0\in[1.866,2.081]$ \\ \hline
\end{tabular}%
\\[0pt]
\end{center}
\caption{Collision types versus the magnitude of the initial kick, $k_{0}$.}
\label{domainecollisionk0solfond}
\end{table}

\begin{figure}[th]
\centering
\subfigure[]{\label{absusolfond1693}%
\includegraphics[height=3.5cm]{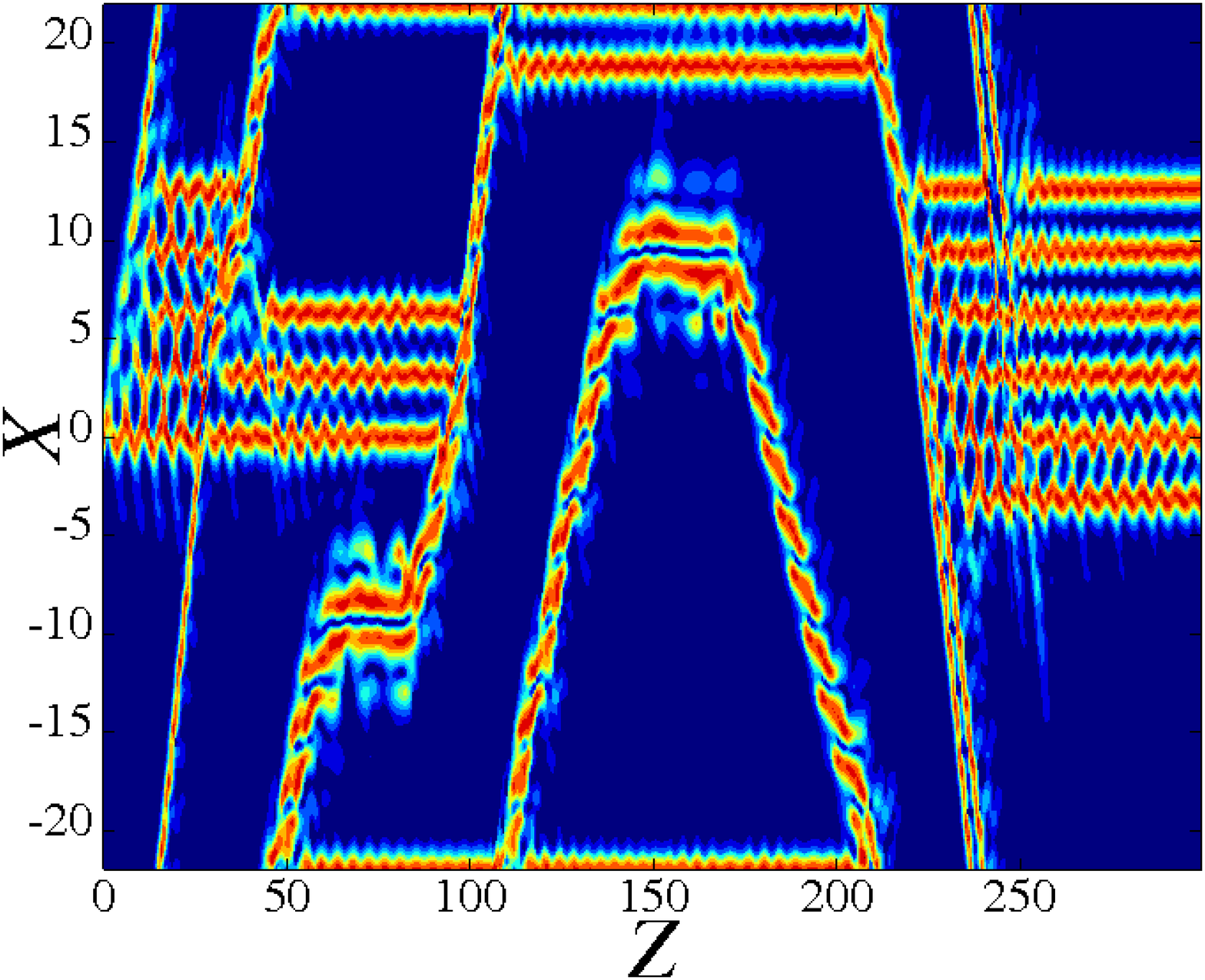}}%
\subfigure[]{\label{energsolfond1693coll}%
\includegraphics[height=3.5cm]{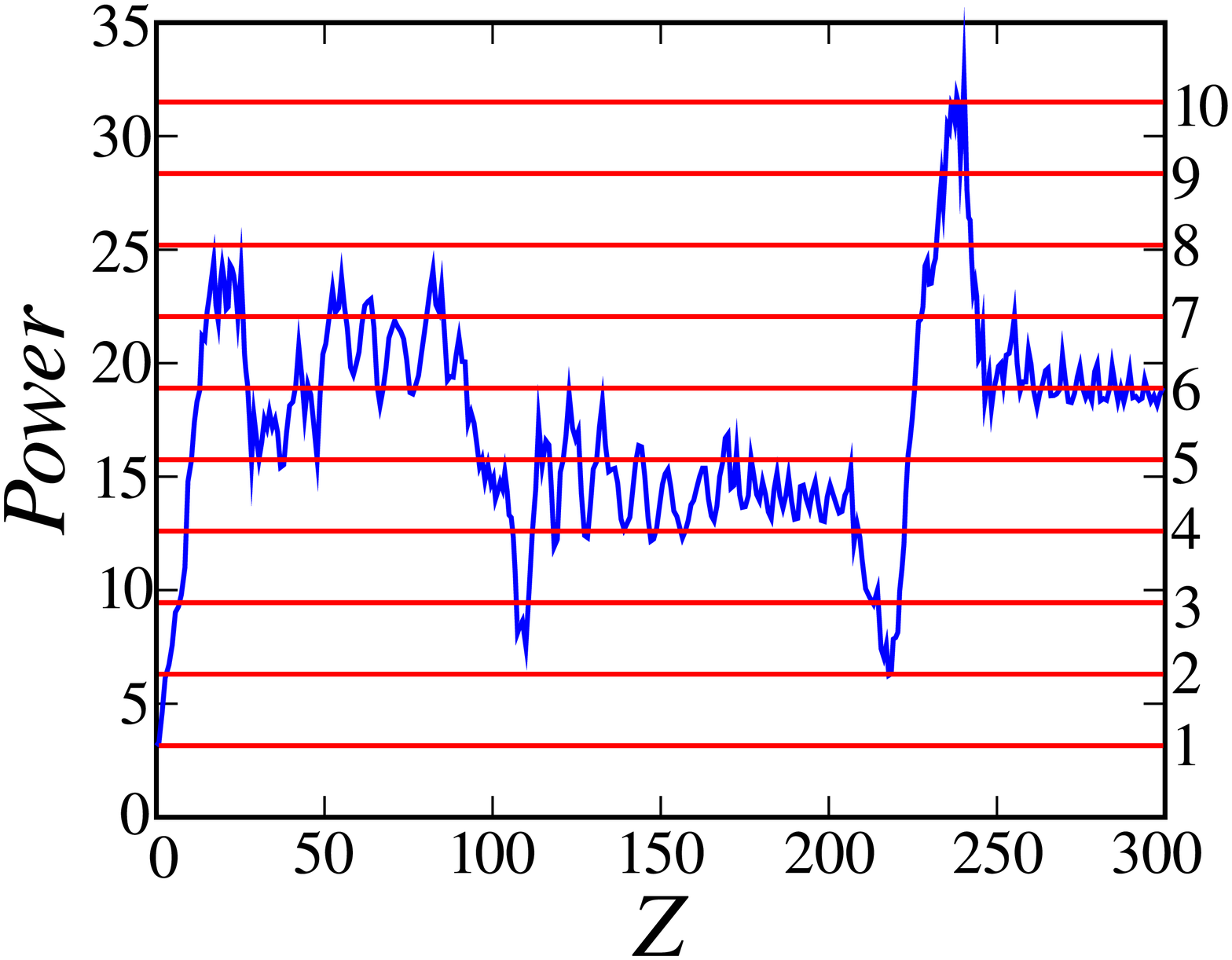}}
\caption{(Color online) (a): An example of the complex interaction of the
free soliton with the standing structure at $k_{0}=1.693$ and $\protect%
\theta =0$. (a) The distribution of the field, $\left\vert u\left(
X,Z\right) \right\vert $, in the cross section $Y=0$; (b) the total power
versus $Z$. In panel (b), horizontal lines show the multiple powers of the
quiescent fundamental soliton. }
\label{solfond1693}
\end{figure}

A special case is the one corresponding to the creation of the largest
number of solitons at $k_{0}=1.693$, as shown above. The collision pattern
is quite complex in this case, as shown in Fig. \ref{absusolfond1693}. Both
elastic collisions (at $Z\simeq 50$ and $100$) and absorptions (at $Z\simeq
238$) are observed. An unexpected feature of the process is the reversal of
the direction of motion of two solitons around $Z\simeq 150$. The whole
patterns eventually relaxes into an array built of six solitons, which is
confirmed by the power-evolution plot in Fig. \ref{energsolfond1693coll}.

\begin{figure}[th]
\centering
\subfigure[]{\label{vitsolfond1974theta0}%
\includegraphics[height=3.5cm]{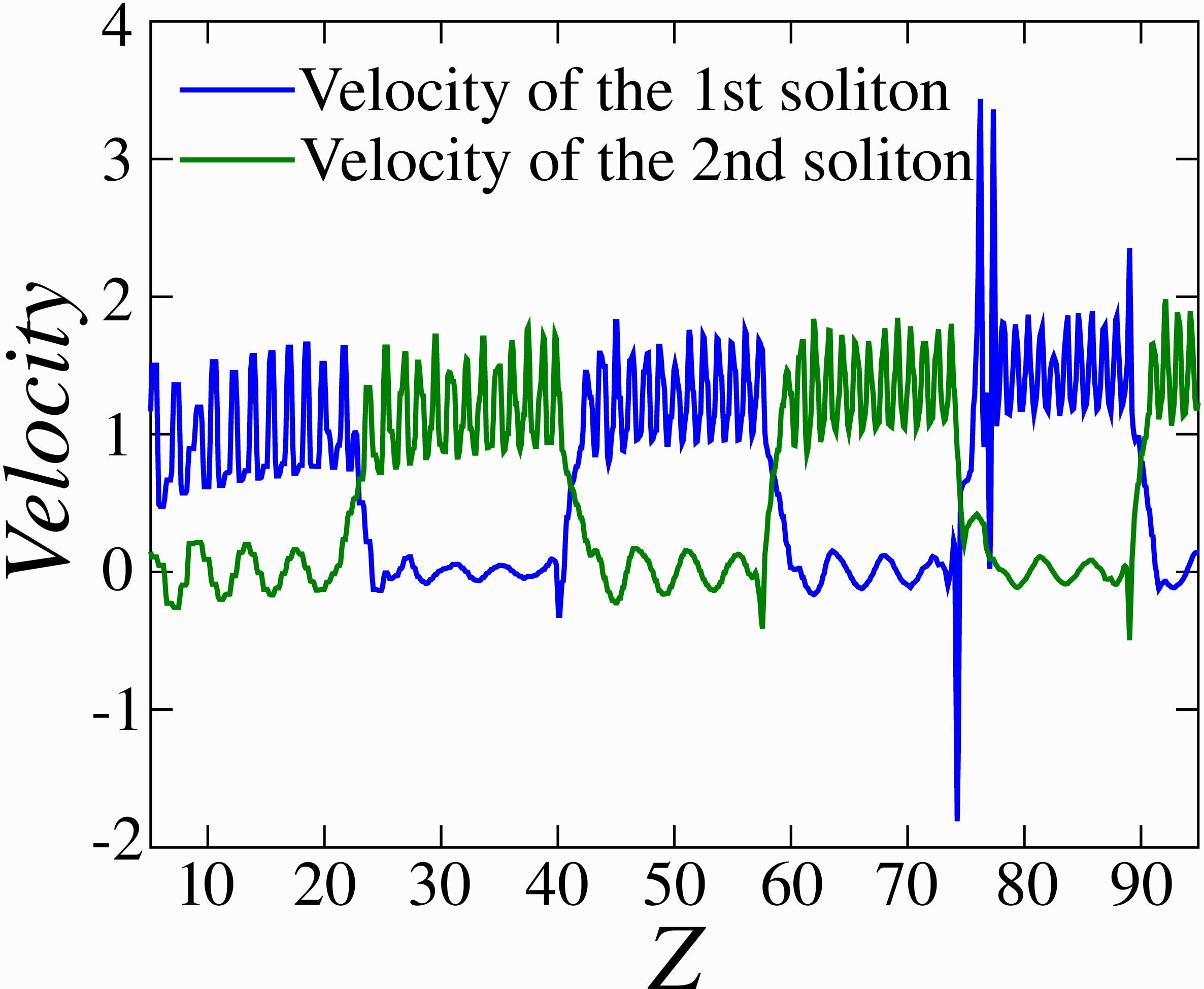}} %
\subfigure[]{\label{vitsolfond1974theta0complet}%
\includegraphics[height=3.5cm]{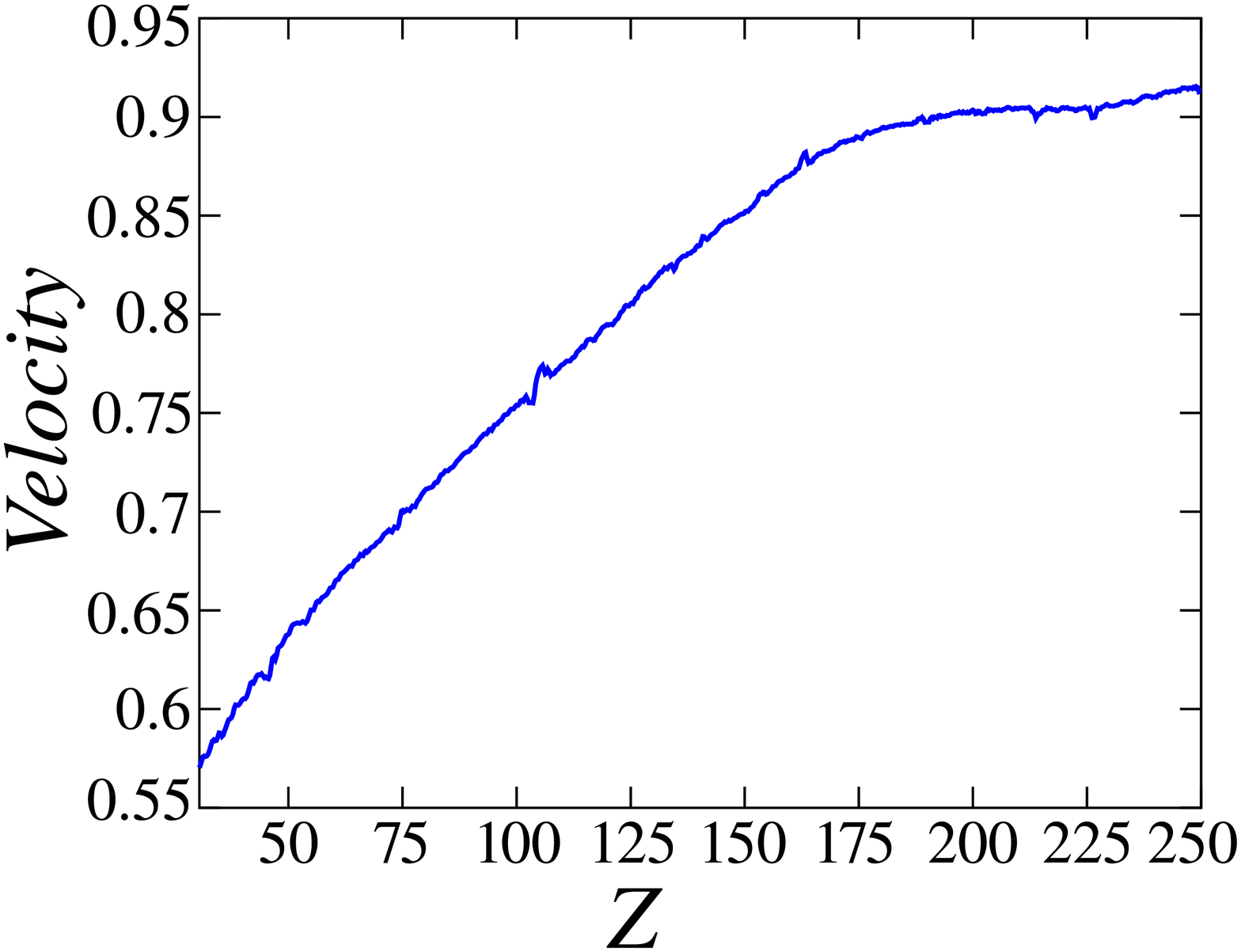}}
\caption{(Color online) Velocities of each soliton (a) and center of mass
(b) versus $Z$ for the pair of periodically colliding fundamental solitons.
Here $k_{0}=1.974$ and $\protect\theta =0$.}
\label{solfond1974theta0}
\end{figure}

In the case of elastic collisions between two solitons recurring
indefinitely long, we have checked if the velocity of the transmitted
soliton is the same as that of the incident one. To this end, it is relevant
to consider the case of $k_{0}=1.974$ and $\theta =0$. After the first round
trip of the emitted soliton, the one original is quiescent while the emitted
one is running into it.
It is seen in Fig. \ref{vitsolfond1974theta0} that each collision leads to
the exchange of velocities between the two solitons, as for colliding hard
particles. In addition, we identified the velocity of the center of mass of
the two-solitons set. Figure \ref{vitsolfond1974theta0complet} shows that
the latter velocity gradually increases within the sequence of collisions,
approaching one of the above-mentioned discrete values characteristic to the
established motion of single solitons. In the present case, the asymptotic
velocity is close to $1$
. 

\section{Conclusions}

The subject of this work is the mobility of 2D dissipative solitons in the
CGL (complex Ginzburg-Landau) equation which includes the spatially periodic
potential. This equation models bulk lasing media with built-in transverse
gratings. The soliton was set in motion by the application of the kick,
which corresponds to a tilt of the seed beam \cite{Vladimirov}. The mobility
implies the possibility to generate oblique laser beams in the medium.
Further, the advancement of the kicked soliton may be used for the
controllable creation of various arrayed patterns in the wake of the soliton
hopping between the potential cells. The depinning threshold, i.e., the
smallest strength of the kick which sets the quiescent soliton into motion,
was found by means of simulations, and also with the help of the analytical
approximation, based on the estimate of the condition for the passage of the
kicked object across the PN (Peierls-Nabarro) potential. The dependence of
the threshold on the orientation of the kick with respect to the underlying
lattice potential was studied too. Various pattern-formation scenarios have
been identified above the threshold, with the number of solitons in
stationary arrayed patterns varying from one to six. Freely moving solitons
may eventually assume two distinct values of the velocity, which represent
coexisting attractors in this dissipative system. Also studied were elastic
and inelastic collisions between the free soliton and stationary
multi-soliton structures, with two generic outcomes: the quasi-elastic
passage, like in the case of the Newton's cradle, and absorption of the free
soliton by the quiescent structure (sometimes, after several passages). A
natural extension of the present work may deal with the dynamics initiated
by the application of the kick to vortices pinned by the underlying grating.

\section*{Acknowledgments}

This work was supported, in a part, by grant No. 3-6738 from High Council
for scientific and technological cooperation between France and Israel. The
work of DM was supported in part by a Senior Chair Grant from R{\'{e}}gion
Pays de Loire, France. Support from the Romanian Ministry of Education and
Research (Project PN-II-ID-PCE-2011-3-0083) is also acknowledged.

\end{document}